\documentclass[12pt]{article}
\usepackage{amsmath}
\usepackage{amssymb}
\usepackage{amsthm}
\usepackage{graphicx}
\usepackage{enumerate}
\usepackage{natbib}
\usepackage{url} 
\usepackage{enumitem}

\newcommand{\blind}{1}

\newcommand{\bx}{\mathbf{x}}

\newcommand{\bs}{\mathbf{s}}
\newcommand{\bX}{\mathbf{X}}

\newcommand{\bh}{\mathbf{h}}
\newcommand{\by}{\mathbf{y}}

\newcommand{\br}{\mathbf{r}}

\newcommand{\bchi}{\boldsymbol{\chi}}
\newcommand{\bEta}{\boldsymbol{\eta}}

\newcommand{\bmu}{\boldsymbol{\mu}}
\newcommand{\btheta}{\boldsymbol{\theta}}

\newcommand{\bdelta}{\boldsymbol{\delta}}

\newcommand{\brho}{\boldsymbol{\rho}}

\newcommand{\bzero}{\mathbf{0}}
\newcommand{\bpsi}{\boldsymbol{\psi}}

\def\T{{ T }}

\begin{document}

\def\spacingset#1{\renewcommand{\baselinestretch}%
{#1}\small\normalsize} \spacingset{1}


\if1\blind
{
  \title{\bf General Bayesian $L^2$ calibration of mathematical models}
  \author{Antony M. Overstall \\
    School of Mathematical Sciences, University of Southampton,\\ Southampton SO17 1BJ, U.K., A.M.Overstall@soton.ac.uk\\
    and \\
    James M. McGree \\
    School of Mathematical Sciences, Queensland University of Technology,\\ Brisbane 4001, Australia, james.mcgree@qut.edu.au
    }
  \maketitle
} \fi

\if0\blind
{
  \bigskip
  \bigskip
  \bigskip
  \begin{center}
    {\LARGE\bf General Bayesian $L^2$ calibration of mathematical models}
\end{center}
  \medskip
} \fi

\bigskip
\begin{abstract}
A mathematical model is a function taking certain arguments and returning a theoretical prediction of a feature of a physical system. The arguments to the mathematical model can be split into two groups; (a) controllable variables of the system; and (b) calibration parameters: unknown characteristics of the physical system that cannot be controlled or directly measured. Of interest is the estimation of the calibration parameter using physical observations. Since the mathematical model will be an inexact representation of the physical system: the aim is to estimate values for the calibration parameters to make the mathematical model ``close" to the physical system. Closeness is defined as the squared $L^2$ norm of the difference between the mathematical model and the physical system. Different Bayesian and general Bayesian methods are introduced, developed and compared for this task.
\end{abstract}

\noindent%
\vfill

\newpage
\spacingset{1.9} 

\section{Introduction} \label{sec:intro}

A mathematical model is a representation of a physical system, often underpinned by scientific theory, which is used to understand, predict and control the physical system. When such models are evaluated by complex, computationally expensive code, they are known as computer models \citep[e.g. ][Section 1.2]{gramacy2020surrogates}.

A mathematical model is a function taking certain arguments and returning a theoretical prediction of a feature of the physical system. Following \cite{plumlee_2017}, the arguments to the mathematical model can be split into two groups; (a) \emph{inputs}: controllable or measurable variables of the system; and (b) \emph{calibration parameters}: unknown characteristics of the physical system that cannot be controlled or directly measured.

This paper addresses calibration: the task of attributing values to the calibration parameters using observations of the physical system. The values given to the calibration parameters should, in some sense, result in the mathematical model (when considered solely as a function of the inputs) being ``close'' to the physical system. This goal should recognise that the mathematical model is an inexact representation of the physical system, i.e. there do not exist values of the calibration parameters such that the mathematical model is equal to the physical system for all values of the inputs. 

In their seminal work, \cite{kennedyohagan2001} considered the bias function, i.e. the difference between the physical system and the mathematical model. Since the bias function is unknown, a Gaussian process prior distribution is assumed for this function. A Bayesian approach is then adopted with the goal of evaluating the marginal posterior distribution of the calibration parameters. However, for the \citeauthor{kennedyohagan2001} approach, the actual parameter values being estimated (termed in this paper: target parameter values) depend on the choice of the covariance function for the Gaussian process prior of the bias function \citep[e.g.][]{tuo_wu_2015a}.

To address this problem, \cite{tuo_wu_2015b} proposed an alternative (frequentist) $L^2$ calibration framework by defining $L^2$ target parameter values as those that minimise the squared norm of the difference between the true physical system and the mathematical model (in the associated $L^2$ space). This is accomplished by minimising a loss function which estimates the squared norm. The specification of this loss makes limited assumptions about the data-generating process. This choice of target parameter values are appealing as they have interpretable physical meaning and do not depend on any modelling choices (beyond the specification of the mathematical model). This has motivated Bayesian analogues of $L^2$ calibration \citep[e.g.][]{plumlee_2017, gu_wang_2018, xie_xu_2020}. However, these approaches can be computationally expensive and make rigid assumptions about the data-generating process. 

The purpose of this paper is twofold. Firstly, to introduce a general Bayesian inference \citep[e.g.][]{bissiri_etal_2016} framework for $L^2$ calibration of mathematical models. Under this approach, unlike traditional Bayesian inference, a so-called generalised posterior distribution for the calibration parameters can be formed using frequentist loss functions without specifying a probabilistic data-generating process for the observations of the physical system. The advantages of the general Bayesian framework, when compared to existing traditional Bayesian calibration approaches, are that it is conceptually and computationally simpler, allows more transparent incorporation of prior information, and, as stated above, does not require specification of a probabilistic data-generating process for the observations (and thus should be less sensitive to misspecification of said data-generating process). The disadvantage is that the scale of the generalised posterior is arbitrary. To address this, we develop an automatic scaling approach adapted from the composite likelihood literature. 

Secondly, it will present a comparison of (general) Bayesian $L^2$ calibration methods, i.e. those that target the $L^2$ calibration parameters, using simulation studies. The methods will be compared in terms of bias (in relation to estimating the $L^2$ target parameter values), uncertainty and coverage of probability intervals. 

The remainder of this paper is organised as follows. In Section~\ref{sec:back}, we provide a background, including a statement of the problem as well as reviews of frequentist and Bayesian approaches to $L^2$ calibration of mathematical models. In Section~\ref{sec:methods}, we introduce general Bayesian $L^2$ calibration including the automatic scaling. Section~\ref{sec:sims} provides a simulation study comparing (general) Bayesian $L^2$ calibration approaches and Section~\ref{sec:balls} we implement the methods on real applications.

\section{Background} \label{sec:back}

\subsection{Setup} \label{sec:setup}

Calibration is performed using $n$ observations of the physical system. That is, for $i=1,\dots,n$, a response $y_i$ is observed of the physical system under inputs $\bx_i = \left(x_{i1},\dots,x_{ik}\right)^\T \in \mathcal{X}$. Here $\mathcal{X}$ denotes the $k$-dimensional input space. Similar to \cite{wong_etal_2017}, we assume, perhaps after transformation, that $\mathcal{X} = [0,1]^k$ with $\mathrm{Vol}(\mathcal{X}) = \int_{\mathcal{X}}\mathrm{d}\bx=1$. Let $\by = \left(y_1,\dots,y_n\right)^\T$ be the $n \times 1$ vector of responses and let $X = \left\{ \bx_1, \dots, \bx_n \right\}$ be the design. 

It is assumed that the true data-generating process for $\by$ is
\begin{equation}
y_i = \mu(\bx_i) + e_i
\label{eqn:true}
\end{equation}
for $i=1,\dots,n$, where $\mu(\bx)$ is an unknown function giving the true value of the physical system at inputs $\bx$, and $e_1,\dots,e_n$ are independent and identically distributed random errors with $\mathrm{E}\left(e_i\right) = 0$ and $\mathrm{var}\left(e_i\right) = \sigma_0^2 < \infty$.

Let $\eta(\bx,\btheta)$ denote the mathematical model where $\btheta = \left(\theta_1,\dots,\theta_p\right)^\T \in \Theta \subset \mathbb{R}^p$ denotes the $p \times 1$ vector of unknown calibration parameters. The purpose of calibration is to use the data, $\by$ and $\bX$, to estimate the calibration parameters. The challenge of doing so is that the mathematical model is inexact, i.e. there do not exist values of the calibration parameters $\btheta_0 \in \Theta$ such that $\eta(\bx, \btheta_0) = \mu(\bx)$ for all $\bx \in \mathcal{X}$. Instead, calibration aims to estimate $\btheta_C \in \Theta$ such that $\eta(\bx, \btheta_C)$ is ``close" to $\mu(\bx)$.

\subsection{$L^2$ calibration} \label{sec:L2calib}

In this paper, we focus on where ``closeness" is defined by the squared norm of the difference between the true physical system and the mathematical model (in the associated $L^2$ space) given by
$$\int_{\mathcal{X}} \left[ \mu(\bx) - \eta(\bx; \btheta) \right]^2 \mathrm{d}\bx.$$
In practice, even if $\mu(\bx)$ was known, the above expression will not be available in closed form. Therefore, we consider the quadrature approximation given by
\begin{equation}
L_{L^2}(\btheta) = \sum_{q=1}^Q \omega_q \left[ \mu(\bchi_q) - \eta(\bchi_q; \btheta) \right]^2,
\label{eqn:L2quad}
\end{equation}
where $\left\{ \bchi_q, \omega_q \right\}_{q=1}^Q$ are quadrature nodes and weights, respectively, and define
$$\btheta_C = \btheta_{L^2} = \arg \min_{\btheta \in \Theta} L_{L^2}(\btheta).$$
Of course, many other choices could be made to define $\btheta_C$. However, $L_{L^2}(\btheta)$ has an appealing physical interpretation.

Under a frequentist approach, \citet{tuo_wu_2015b} define estimators, $\hat{\btheta}_{L^2}$, of $\btheta_{L^2}$ to be
\begin{equation}
\hat{\btheta}_{L^2} = \arg \min_{\btheta \in \Theta} \ell_{L^2}(\btheta; \by),
\label{eqn:hatthetaL2}
\end{equation}
where
\begin{equation}
\ell_{L^2}(\btheta; \by) = \sum_{q=1}^Q \omega_q \left[ \hat{\mu}(\bchi_q; \by, X) - \eta(\bchi_q,\btheta)\right]^2,  \label{eqn:L2loss}
\end{equation}
is termed the \emph{$L^2$ loss} and $\hat{\mu}(\bx; \by)$ is a non-parametric predictor of $\mu(\bx)$ formed from the observations of the physical system. Note that \citet{tuo_wu_2015b} defined the $L^2$ loss with the weighted sum in (\ref{eqn:L2loss}) replaced by a integral over $\mathcal{X}$. We have used the weighted sum as the quadrature approximation to allow practical evaluation.

For the non-parametric predictor, let $c(\cdot,\cdot; \bpsi)$ be a correlation function depending on parameters $\bpsi$. For example, throughout this paper, we use the squared exponential correlation function given by $c(\bx,\bx'; \bpsi) = \exp \left[ - \sum_{j=1}^k \psi_j(x_{j} - x_{j}')^2 \right]$, where $\bpsi = (\psi_1,\dots,\psi_k)^T$ with $\psi_j>0$ for $j=1,\dots,k$. Let $C_{XX}(\bpsi)$ and $\Phi(\bpsi,\kappa)$ be $n \times n$ matrices where $C_{XX}(\bpsi)$ has $ij$th element $c(\bx_i,\bx_j; \bpsi)$ and $\Phi(\bpsi,\kappa) = \kappa I_n + C_{XX}(\bpsi)$, for $\kappa>0$. Then
\begin{equation}
\hat{\mu}(\bx; \by) = \bs_{X}(\bx; \bpsi)^T \Phi(\bpsi, \kappa)^{-1} \by
\label{eqn:muhat}
\end{equation}
where $\bs_X(\bx;\bpsi)$ is an $n \times 1$ vector with $i$th element $c(\bx,\bx_i;\bpsi)$. The parameters $\brho = (\bpsi^T,\kappa)^T$ are estimated by generalised cross-validation, i.e. $$\hat{\brho} = \arg \min  \frac{\by^\T \left[I_n - R(\brho)\right]^2 \by}{\left\{1 - \mathrm{tr}[R(\brho)]/n\right\}^2},$$ where $R(\brho) = C_{XX}(\bpsi)\Phi(\bpsi,\kappa)^{-1}$. 

Under certain conditions, one of the most stringent of which is that the elements, $\left\{\bx_1,\dots,\bx_n\right\}$, of the design $X$ are realisations of independent and identically distributed random variables from the uniform distribution over $\mathcal{X}$, \citet{tuo_wu_2015b} show that $\hat{\btheta}_{L^2}$ are consistent estimators of $\btheta_{L^2}$ and have an asymptotic normal distribution.


Alternatively, \citet{tuo_wu_2015b} considered the OLS estimator given by
$$
\hat{\btheta}_{OLS} = \mathrm{arg} \min_{\btheta \in \Theta} \ell_{OLS}(\btheta; \by),
$$
where
\begin{equation}
\ell_{OLS}(\btheta; \by) = \sum_{i=1}^n \left[y_i - \eta(\bx_i, \btheta) \right]^2
\label{eqn:OLSloss}
\end{equation}
is the \emph{OLS loss}. Under the same requirement that the elements of $X$ be a random sample from the uniform distribution over $\mathcal{X}$, \citet{tuo_wu_2015b} show that $\hat{\btheta}_{OLS}$ are also consistent estimators of $\btheta_{L^2}$ and also have an asymptotic normal distribution. \citet{tuo_wu_2015b} observed that the asymptotic variance matrix of $\hat{\btheta}_{OLS}$ is greater or equal to that of $\hat{\btheta}_{L^2}$ (in the L\"{o}wner ordering sense). Moreover, equality is only achieved when there exist $\btheta_0$ such that $\mu(\bx) = \eta(\bx, \btheta_0)$ for all $\bx \in \mathcal{X}$, i.e. the mathematical model is exact. \cite{wong_etal_2017} studied frequentist OLS calibration under a fixed, non-random design, and, under certain conditions, showed that the resulting $\hat{\btheta}_{OLS}$ are consistent estimators of $\btheta_{L^2}$. 

\subsection{Bayesian $L^2$ calibration} \label{sec:bayesL2}

Bayesian approaches allow a coherent approach to uncertainty quantification. Due to this, several authors have proposed Bayesian $L^2$ calibration approaches, and in Section~\ref{sec:sims}, we compare these with general Bayesian $L^2$ calibration approaches, which we introduce in Section~\ref{sec:methods}. The methods we consider are Bayesian non-linear regression (Section~\ref{sec:bnlm}), a modified version of \cite{kennedyohagan2001} (Section~\ref{sec:koh}) and a projected $L^2$ approach (Section~\ref{sec:proj}). 

\subsubsection{Bayesian non-linear regression} \label{sec:bnlm}

\cite{walker_2013} showed that Bayesian inference under a misspecified model actually targets the parameter values that minimise the Kullback-Liebler divergence between the true probability distribution for the responses (in our case, given by (\ref{eqn:true})) and that imposed by the assumed probabilistic model. To demonstrate Bayesian inference under a misspecified mathematical model, consider a simple Bayesian approach to calibrating the mathematical model, i.e. a Bayesian non-linear regression model. It is assumed, incorrectly, that
\begin{equation}
y_i \sim \mathrm{N} \left[ \eta(\bx_i, \btheta_0) , \sigma^2_0 \right]
\label{eqn:BNLM}
\end{equation}
independently, for $i=1,\dots,n$. That is, it has been assumed that there do exist $\btheta_0$ such that $\eta(\bx, \btheta_0) = \mu(\bx)$ for all $\bx \in \mathcal{X}$ (or, at least for the elements of the design $X$), that the errors are normally distributed.

However, in reality the mathematical model is inexact. In this case, the result of \cite{walker_2013} allows us to find the target parameter values, labelled  here $\btheta_{NLM}$ and $\sigma^2_{NLM}$. In Section~\ref{sec:bnlm2} in the Supplementary Material, we show that if the elements of the design $X = \left\{ \bx_1, \dots, \bx_n \right\}$ are uniformly-generated, or have properties of being uniformly-generated, then the target parameter values, $\btheta_{NLM}$, are approximately $\btheta_{L^2}$.  We also show the target parameter value for the error variance is
\begin{equation}
\sigma^2_{NLM} = \sigma_0^2 + \frac{1}{n} \sum_{i=1}^n \left[ \mu(\bx_i) - \eta(\bx_i, \btheta_{NLM}\right]^2.
\label{eqn:nlmlambda}
\end{equation}
This means, under a suitable design $X$, Bayesian non-linear regression approximately targets $\btheta_{L^2}$. However, observe from (\ref{eqn:nlmlambda}) that the target error variance is $\sigma_{NLM} > \sigma_0^2$, unless the model is true. This overestimation has implications on uncertainty quantification of $\btheta_{NLM} \approx \btheta_{L^2}$ derived from the resulting posterior distribution, e.g. probability intervals for elements of $\btheta_{NLM}$ have coverage larger than the nominal value. We demonstrate this using the simulation studies in Section~\ref{sec:sims}.

For completeness, the posterior distribution of $\btheta_{NLM}$ and $\sigma^2_{NLM}$ is given by
\begin{equation}
\pi_{NLM}(\btheta, \sigma^2 \vert \by) \propto  \exp \left\{ - \frac{\sum_{i=1}^n \left[ \mu(\bx_i) - \eta(\bx_i, \btheta\right]^2}{2 \sigma^2} \right\} \pi_{NLM}(\btheta,\sigma^2),
\label{eqn:post4bnlm}
\end{equation}
where $\pi_{NLM}(\btheta,\sigma^2)$ gives the joint prior distribution for $\btheta_{NLM}$ and $\sigma^2_{NLM}$.

\subsubsection{Modified \citeauthor{kennedyohagan2001} calibration} \label{sec:koh}

First, we need to introduce the original \citet{kennedyohagan2001} framework. It is assumed that the $i$th observational error has $e_i \sim \mathrm{N}(0,\sigma^2_0)$, for $i=1,\dots,n$, and
\begin{equation}
\mu(\bx) = \eta(\bx, \btheta_0) + \delta_0(\bx),
\label{eqn:koh}
\end{equation}
where $\delta_0(\bx)$ is an unknown bias function giving the difference between the true physical system and the mathematical model (evaluated at $\btheta_0$). \citet{kennedyohagan2001} imposed a zero-mean Gaussian process prior distribution for the bias function, i.e. $\delta_0(\cdot) \sim \mathrm{GP}\left[0,\sigma^2_0 c(\cdot,\cdot;\bpsi_0)/\kappa_0\right]$ where $c(\cdot,\cdot;\cdot)$ is the correlation function as defined in Section~\ref{sec:L2calib}. A fully Bayesian approach is then taken by evaluating the joint posterior distribution of the unknown parameters (using, for example, MCMC), following specification of a joint prior distribution.

A key question is: what values of the calibration parameters, here labelled $\btheta_{KOH}$, are actually being targeted? The Kullback-Liebler divergence between the true probability distribution for the responses, and that imposed by the assumed probabilistic model, is minimised by equivalently minimising
\begin{eqnarray*}
L_{KOH}(\btheta,\sigma^2, \kappa, \bpsi) & = & \frac{1}{2} \log \vert \Sigma (\sigma^2, \kappa, \bpsi) \vert + \frac{1}{2} \left[ \bmu_X - \bEta_X(\btheta)\right]^T \Sigma(\sigma^2, \kappa, \bpsi)^{-1} \left[ \bmu_X - \bEta_X(\btheta)\right]^T\\
& & \mbox{  } \qquad  + \frac{\sigma_0^2}{2}\mathrm{tr} \left[ \Sigma(\sigma^2, \kappa, \bpsi)^{-1} \right],
\end{eqnarray*}
with respect to $\btheta$, $\sigma^2$, $\kappa$ and $\bpsi$, where $\bmu_X = \left[ \mu(\bx_1), \dots, \mu(\bx_n) \right]^T$, $\bEta_X(\btheta) = \left[ \eta(\bx_1,\btheta), \dots, \bEta(\bx_n,\btheta) \right]^T$, and 
$\Sigma(\sigma^2, \kappa, \bpsi) = \sigma^2 + \sigma^2C_{XX}(\bpsi)/\kappa=\sigma^2  \Phi(\bpsi,\kappa)/\kappa$, with $C_{XX}(\bpsi)$ and $\Phi(\bpsi,\kappa)$ as defined in Section~\ref{sec:L2calib}. The target parameter values $\btheta_{KOH}$ that minimise $L_{KOH}(\btheta,\sigma^2, \kappa, \bpsi)$ will depend on the choice of correlation function $c(\cdot,\cdot;\cdot)$; a criticism of the \cite{kennedyohagan2001} framework made by several authors \citep[e.g.][]{tuo_wu_2015b,tuo_wu_2015a, plumlee_2017, wong_etal_2017}. 

Instead, \cite{plumlee_2017} developed a modification of the \citeauthor{kennedyohagan2001} framework with the aim to target $\btheta_{L^2}$. Recall the definition of $\btheta_{L^2}$ as those values of $\btheta$ that minimise $L_{L^2}(\btheta)$. Then $\mathrm{d} L_{L^2}(\btheta_{L^2})/\mathrm{d}\btheta = \bzero_p$, where
$$\frac{\mathrm{d} L_{L^2}(\btheta)}{\mathrm{d}\btheta} = -2 \int_{\mathcal{X}} \frac{\partial \eta(\bx; \btheta)}{\partial \btheta} \left[ \mu(\bx) - \eta(\btheta;\bx)\right] \mathrm{d}\bx.$$
Let the bias function be $\delta_{L^2}(\bx) = \mu(\bx) - \eta(\btheta_{L^2};\bx)$, then $\bzero_p = -2 \int_{\mathcal{X}} \frac{\partial \eta(\bx; \btheta_{L^2})}{\partial \btheta} \delta_{L^2}(\bx) \mathrm{d}\bx$, i.e. $\delta_{L^2}(\bx)$ is orthogonal to $\partial \eta(\bx; \btheta_{L^2})/\partial \btheta$. \cite{plumlee_2017} modified the correlation function of the Gaussian process prior for the bias function to impose this orthogonality. 

In particular, the modified correlation function is $c_P(\bx,\bx'; \btheta_{L^2}, \bpsi)$, where
\begin{equation}
c_P(\bx,\bx'; \btheta, \bpsi) =c(\bx,\bx'; \bpsi) - \bh(\bx; \btheta, \bpsi)^T H(\btheta,\bpsi)^{-1}\bh(\bx'; \btheta, \bpsi),
\label{eqn:KOHcorr}
\end{equation}
where $\bh(\bx; \btheta, \bpsi)$ is a $p \times 1$ vector with $j$th element
\begin{equation}
\sum_{q=1}^Q \omega_q \frac{\partial \eta(\bchi_q;\btheta)}{\partial \btheta} c(\bx, \bchi_q;\bpsi),
\label{eqn:h}
\end{equation}
and $H(\btheta, \bpsi)$ is a $p \times p$ matrix with $jl$th element 
\begin{equation}
\sum_{q_1=1}^Q \sum_{q_2=1}^Q\omega_{q_1}\omega_{q_2} \frac{\partial \eta(\bchi_{q_1};\btheta)}{\partial \btheta}\frac{\partial \eta(\bchi_{q_2};\btheta)}{\partial \btheta}c(\bchi_{q_1}, \bchi_{q_2};\bpsi).
\label{eqn:H}
\end{equation}

Note that, \cite{plumlee_2017} originally defined the modified correlation function with $\btheta_{L^2}$ replaced by $\btheta$ in the evaluation of $\partial \eta(\cdot;\btheta)/\partial \btheta$ in (\ref{eqn:h}) and (\ref{eqn:H}), resulting in an modified correlation function depending on $\btheta$. We have made our modification so that it is more straightforward to show that the target calibration parameter values, denoted $\btheta_{PKOH}$, are approximately equal to $\btheta_{L^2}$ if the elements of $X$ are uniformly-generated, or have properties of being uniformly-generated. See Section~\ref{sec:mkoh2}, in the Supplementary Material, for a justification. In practice, we evaluate the modified correlation function by replacing $\btheta_{L^2}$ in (\ref{eqn:KOHcorr}) by $\hat{\btheta}_{L^2}$. 

%

\subsubsection{Projected $L^2$ calibration} \label{sec:proj}

An alternative projected Bayesian $L^2$ approach proposed by \citet{xie_xu_2020} begins by assuming a zero mean Gaussian process for $\mu(\bx)$, i.e. $\mu(\cdot) \sim \mathrm{GP}\left[0, \gamma_0(\cdot, \cdot; \bpsi_0)\right]$. The resulting posterior distribution for $\mu(\bx)$ is a normal distribution with mean $\hat{\mu}(\bx; \by)$ given by (\ref{eqn:muhat}) and variance $\nu(\bx; \by) = \gamma \left[c(\bx,\bx; \bpsi) - \bs_X(\bx; \bpsi)^T \Phi(\bpsi)^{-1} \bs_X(\bx;\bpsi)\right]$. By writing $L_{L^2}(\btheta)$ as a functional of $\mu(\cdot)$, i.e.
\begin{equation}
L_{L^2}\left[\btheta, \mu(\cdot)\right] = \sum_{q=1}^Q \omega_q \left[\mu(\bchi_q) - \eta(\btheta,\bchi_q) \right]^2,
\label{eqn:xiexi}
\end{equation}
\citet{xie_xu_2020} noted that this induces a posterior distribution for $L_{L^2}\left[\btheta, \mu(\cdot)\right]$ and hence for the values of $\btheta$ minimizing $L_{L^2}\left[\btheta, \mu(\cdot)\right]$. 

In Section~\ref{sec:xx2} of the Supplementary Material, we show that the target calibration parameter values, $\btheta_{PL}$ are approximately $\btheta_{L^2}$.

In practice, a Monte Carlo sample from the posterior distribution of the calibration parameters can be generated in a straightforward manner. First, sample from the posterior distribution of $\mu(\bx)$ evaluated at the quadrature points $\bchi_1,\dots,\bchi_Q$ (essentially, generating from a $Q$-variate normal distribution) and then, for each sampled function values, minimise $L_{L^2}\left[\btheta, \mu(\cdot)\right]$.

\section{General Bayesian calibration of mathematical models} \label{sec:methods}

\subsection{General Bayesian inference} \label{sec:gbi}

We now provide a brief outline of general Bayesian inference for calibrating a mathematical model. Typically, general Bayesian inference begins with the specification of a loss function denoted $\ell(\btheta; \by)$. This function identifies desirable values for the calibration parameters based on observations $\by$. \cite{bissiri_etal_2016} showed that general Bayesian inference provides coherent inference about the target parameter values: $\btheta_{\ell} = \arg \min_{\btheta \in \Theta} L(\btheta)$ where $L_{\ell}(\btheta) = \mathrm{E}_{\by}\left[ \ell(\btheta; \by) \right]$ is the expected loss under the true probability distribution of the observations $\by$.

General Bayesian inference proceeds via the generalised (or Gibbs) posterior distribution given by
\begin{equation}
\pi_{\ell}(\btheta \vert \by) \propto \exp \left[ - \ell(\btheta; \by) \right] \pi_{\ell}(\btheta),
\label{eqn:genpost}
\end{equation}
where $\pi_{\ell}(\by | \btheta) = \exp \left[ - \ell(\btheta; \by) \right]$ is known as the \emph{generalised likelihood} and $\pi_{\ell}(\btheta)$ is the probability density function (pdf) of the prior distribution for $\btheta_{\ell}$. 

The traditional Bayesian posterior distribution can be viewed as a general Bayesian posterior under the \emph{self-information loss} $\ell_{SI}(\btheta; \by) = - \log \pi(\by \vert \btheta)$, where $\pi(\by | \btheta)$ is the likelihood function which follows from specification of a probabilistic model for the observations $\by$. We can recover the result of \cite{walker_2013} from Section~\ref{sec:bayesL2}, by noting that under the self-information loss, the target parameters $\btheta_{SI}$ are those values of $\btheta$ that minimise 
$$L_{SI}(\btheta) = \mathrm{E}_{\by}\left[ \ell_{SI}(\btheta; \by)\right] = - \mathrm{E}_{\by}\left[\log \pi(\by \vert \btheta)\right].$$
It follows that $\btheta_{SI}$ minimise the Kullback-Liebler divergence between the probabilistic model assumed for $\by$ and the true distribution.

\subsection{General Bayesian $L^2$ calibration of mathematical models} \label{sec:gbicalib}

General Bayesian $L^2$ calibration of mathematical models follows from using $\ell_{L^2}(\btheta)$, as defined in (\ref{eqn:L2loss}), as the loss function. The target parameter values minimise the expected loss given by
$$\sum_{q=1}^Q \omega_q \left\{ \mathrm{E}_{\by}\left[ \hat{\mu}(\bchi_q; \by,X) \right] - \eta(\bchi_q;\btheta) \right\}^2 + \sum_{q=1}^Q \omega_q \mathrm{var}\left[ \hat{\mu}(\bchi_q; \by) \right],$$
with respect to $\btheta$. If $\mathrm{E}\left[ \hat{\mu}(\bchi_q; \by) \right] \approx \mu(\bchi_q)$, for all $q=1,\dots,Q$, then the target parameter values will approximately be equal to $\btheta_{L^2}$. In other words, if $\hat{\mu}(\cdot; \by)$ is an approximately unbiased predictor of $\mu(\bx)$, then the target parameter values will be $\btheta_{L^2}$.

However, a near-universal hurdle to the implementation of general Bayesian inference is that the scale of the general Bayesian posterior is arbitrary. To see, this consider the loss function $\gamma \ell_{L^2}(\btheta)$, where $\gamma>0$, with general Bayesian posterior
$$\pi_{L^2}(\btheta \vert \by) \propto \exp \left[ - \gamma \ell_{L^2}(\btheta; \by)\right] \pi_{L^2}(\btheta).$$
The target parameter values are unaffected by the specification of $\gamma$ since the values of $\btheta$ that minimise the expectation of $\ell_{L^2}(\btheta; \by)$ and $\gamma \ell_{L^2}(\btheta; \by)$ are equal. However, as $\gamma \to 0$, then the general Bayesian posterior converges to the prior distribution for $\btheta_{L^2}$, as given by $\pi_{L^2}(\cdot)$. Conversely, if $\gamma \to \infty$, then the general Bayesian posterior converges to a point mass at $\hat{\btheta}_{L^2}$, as defined in (\ref{eqn:hatthetaL2}). Therefore, $\gamma$ controls the rate of learning from prior to posterior distribution, and its specification is crucial. For example, assuming a non-informative prior distribution for $\btheta_{L^2}$, if $\gamma$ is too small, then the general Bayesian posterior will be too diffuse (for example, in terms of coverage of posterior intervals) and if it is too large, then the general Bayesian posterior will be too concentrated. In Section~\ref{sec:automatic}, we introduce an automatic approach to the specification of $\gamma$.

\subsection{Automatic scaling} \label{sec:automatic}

In this section, we introduce an automatic scaling of the general Bayesian posterior, i.e. the specification of $\gamma$. \cite{woody_etal_2019} considered this problem and devised an approach to automatically specifying $\gamma$ by equating properties of frequentist estimators with properties of the general Bayesian posterior. This was accomplished by using a bootstrapping routine. We also take a similar, but asymptotic and deterministic approach by adapting ideas from the composite likelihood literature \citep{pauli_etal_2011, ribatet_etal_2012}. 

We make the following assumptions. We investigate the veracity of these assumptions, via simulation studies, in Section~\ref{sec:sims}. 

\begin{enumerate}
\item[(i)]
As $n \to \infty$, $\hat{\btheta}_{L^2} \to \btheta_{L^2}$.
\item[(ii)]
The $L^2$ loss can be expressed as
$$\ell_{L^2}(\btheta; \by) = \ell_{L^2}(\hat{\btheta}_{L^2}; \by) + \frac{1}{2} \left( \btheta - \hat{\btheta}_{L^2}\right)^T \frac{\partial^2 \ell_{L^2}(\hat{\btheta}_{L^2}; \by)}{\partial \btheta \partial \btheta^T}\left( \btheta - \hat{\btheta}_{L^2}\right) + r_1(\btheta; \by),$$
with $r_1(\btheta; \by) \to 0$ as $n \to \infty$.
\item[(iii)]
The gradient of the $L^2$ loss can be expressed as
$$\frac{\partial \ell_{L^2}(\btheta; \by)}{\partial \btheta} = \frac{\partial \ell_{L^2}(\btheta_{L^2}; \by)}{\partial \btheta} + \frac{\partial^2 \ell_{L^2}(\btheta_{L^2}; \by)}{\partial \btheta \partial \btheta^T} \left( \btheta - \btheta_{L^2}\right) + \br_2(\btheta; \by),$$
with $\br_2(\btheta; \by) \to \bzero_p$ as $n \to \infty$.
\item[(iv)]
The Hessian of the $L^2$ loss has 
$$\frac{\partial^2 \ell_{L^2}(\btheta_{L^2}; \by)}{\partial \btheta \partial \btheta} \to V = \frac{\partial L^2(\btheta_{L^2}; \by)}{\partial \btheta},$$
as $n \to \infty$
\end{enumerate}

Consider an analogy of the likelihood ratio test statistic
$$
\Lambda(\btheta_{L^2};\by) = 2 \gamma \left[ \ell_{L^2}(\btheta_{L^2}; \by) - \ell_{L^2}(\hat{\btheta}_{L^2}; \by)\right],
$$
i.e. where the log-likelihood has been replaced by the negative $L^2$ loss function. The idea is to specify $\gamma$ so that the asymptotic expectation of $\Lambda(\btheta_{L^2};\by)$ matches $p$: the asymptotic expectation of the likelihood ratio test statistic.

By assumptions (i), (ii) and (iv),
$$\Lambda(\btheta_{L^2};\by) \to \gamma \left( \btheta_{L^2} - \hat{\btheta}_{L^2}\right)^T V \left( \btheta_{L^2} - \hat{\btheta}_{L^2}\right),$$
as $n \to \infty$. Next, by definition and assumption (iii)
\begin{eqnarray*}
\bzero_p & = & \frac{\partial \ell_{L^2}(\hat{\btheta}_{L^2}; \by)}{\partial \btheta}\\
& = & \frac{\partial \ell_{L^2}(\btheta_{L^2}; \by)}{\partial \btheta} + \frac{\partial^2 \ell_{L^2}(\btheta_{L^2}; \by)}{\partial \btheta \partial \btheta^T} \left(\hat{\btheta}_{L^2} - \btheta_{L^2} \right) + \br_2(\hat{\btheta}_{L^2};\by).
\end{eqnarray*}
By rearranging and assumptions (i), (iii) and (iv), $\mathrm{E}_{\by}(\hat{\btheta}_{L^2}) \to \btheta_{L^2}$ and $\mathrm{var}_{\by}(\hat{\btheta}_{L^2}) \to V^{-1} W_{L^2} V^{-1}$, where
\begin{eqnarray*}
W_{L^2} & = & \mathrm{var}_{\by} \left[ \frac{\partial \ell_{L^2}(\btheta_{L^2}; \by)}{\partial \btheta} \right]\\
& = & 4 \sigma_0^2 D \Phi(\bpsi, \kappa)^{-1}\Phi(X; \bpsi, \kappa)^{-1} D^T,
\end{eqnarray*}
with $D$ the $p \times n$ matrix
$$D = \sum_{q=1}^Q \omega_q \frac{\partial \eta(\btheta_{L^2}; \bchi_q)}{\partial \btheta} \bs_X(\bchi_q; \bpsi)^T.$$
Then the asymptotic expectation of $\Lambda(\btheta_{L^2};\by)$ is $\gamma \mathrm{tr} \left( V^{-1} W_{L^2}\right)$, and we specify 
$$\gamma = \frac{p}{\mathrm{tr} \left( V^{-1} W_{L^2}\right)}.$$
In practice we, replace $V$ by $\hat{V} = \frac{\partial^2 \ell_{L^2}(\hat{\btheta_{L^2}}; \by)}{\partial \btheta \partial \btheta}$ and $\sigma_0^2$ by an estimate derived from the non-parametric predictor of $\mu(\bx)$, i.e.
\begin{equation}
\hat{\sigma}_0^2 = \frac{\sum_{i=1}^n \left[ y_i - \hat{\mu}(\bx_i; \by) \right]^2}{\mathrm{tr} \left[ \left(I_n - R(\brho)\right)^2 \right]}.
\label{eqn:sighat}
\end{equation}

Inspired by \cite{woody_etal_2019}, we also consider an approach where the expectation of $\Lambda(\btheta_{L^2};\by)$ is determined via bootstrapping. The details of this are given in Section~\ref{sec:automaticboot} of the Supplementary Material. 

\subsection{General Bayesian OLS calibration of mathematical models} \label{sec:gbiols}

We can also consider a general Bayesian approach using $\ell_{OLS}(\btheta; \by)=\sum_{i=1}^n \left[y_i - \eta(\bx_i;\btheta)\right]^2$, as defined in (\ref{eqn:OLSloss}), as the loss function. Under such a loss function, the target parameter values, were shown in Section~\ref{sec:bnlm} to be approximately equal to $\btheta_{L^2}$. Again noting that the target parameter values are unaffected by multiplying the loss by a positive constant $\gamma$, the general Bayesian posterior is given by
$$\pi_{OLS}(\btheta \vert \by) \propto \exp \left\{ - \gamma \sum_{i=1}^n \left[y_i - \eta(\bx_i;\btheta)\right]^2 \right\} \pi_{OLS}(\btheta).$$
This has the same form as the standard posterior distribution under Bayesian non-linear regression, given by (\ref{eqn:post4bnlm}), with $\gamma = 1/2\sigma^2_0$. The key difference is that under the general Bayesian approach we can use an automatic scaling procedure, similar to Section~\ref{sec:automatic}, to specify a value for $\gamma$, whereas in Bayesian non-linear regression, the response variance is estimated with target parameter value $\sigma^2_{NLM} > \sigma_0^2$. We will compare these two approaches in Section~\ref{sec:sims}. 

The automatic scaling procedure for general Bayesian OLS calibration starts by making the same assumptions (i)-(iv) in Section~\ref{sec:automatic}, but replacing $\ell_{L^2}(\btheta; \by)$ by $\ell_{OLS}(\btheta; \by)$. The same argument as Section~\ref{sec:automatic} is followed with $\Lambda(\btheta_{L^2};\by) = 2 \gamma \left[ \ell_{OLS}(\btheta_{L^2}; \by) - \ell_{OLS}(\hat{\btheta}_{L^2}; \by)\right]$, leading to $\gamma = \frac{p}{\mathrm{tr} \left( V^{-1} W_{OLS}\right)}$, with 
$$W_{OLS} = 4 \sigma_0^2 \sum_{i=1}^n \frac{\partial \eta(\bx_i;\btheta_{L^2})}{\partial \btheta}\frac{\partial \eta(\bx_i;\btheta_{L^2})}{\partial \btheta^T}.$$
Again, $\btheta_{L^2}$ is replaced by the estimator $\hat{\btheta}_{OLS}$ and $\sigma_0^2$ by $\hat{\sigma}_0^2$ given by (\ref{eqn:sighat}). 

Similar to Section~\ref{sec:automatic}, we can also use a bootstrapping approach to specify $\gamma$. Details of this are provided in Section~\ref{sec:automaticboot} in the Supplementary Material. 

\section{Comparison of methods using simulation studies} \label{sec:sims}

\subsection{Introduction} \label{sec:simintro}

There are seven (general) Bayesian calibration approaches where the target parameter values are approximately equal to $\btheta_{L^2}$, as follows:
\begin{enumerate}
\item[(i)]
non-linear regression (Section~\ref{sec:bnlm});
\item[(ii)]
modified \citeauthor{kennedyohagan2001} calibration (Section~\ref{sec:koh});
\item[(iii)]
projected $L^2$ calibration (Section~\ref{sec:proj});
\item[(iv)]
general Bayesian calibration under $L^2$ loss (Section~\ref{sec:gbicalib}) with $\gamma$ specified using asymptotic approach;
\item[(v)]
general Bayesian calibration under OLS loss (Section~\ref{sec:gbiols}) with $\gamma$ specified using asymptotic approach;
\item[(vi)]
general Bayesian calibration under $L^2$ loss (Section~\ref{sec:gbicalib}) with $\gamma$ specified using bootstrap approach;
\item[(vii)]
general Bayesian calibration under OLS loss (Section~\ref{sec:gbiols}) with $\gamma$ specified using bootstrap approach.
\end{enumerate}

We compare these methods using simulation studies. We assume particular configurations for the physical system $\mu(\cdot)$ and mathematical model $\eta(\cdot; \btheta)$. The values $\btheta_{L^2}$ automatically follow from specification of $\mu(\cdot)$ and $\eta(\cdot; \btheta)$. We apply each of the methods listed above and assess their ability in estimating $\btheta_{L^2}$. Specifically, we consider the behaviour of the (generalised) posterior mean, (generalised) posterior standard deviation and the coverage of 95\% probability intervals with respect to the elements of $\btheta_{L^2}$.

\subsection{Configurations} \label{sec:configs}

In each configuration, we consider three different probability distributions for the errors $\epsilon_1,\dots,\epsilon_n$: (i) normal; (ii) t (with $\nu = 3$ degrees of freedom); and (iii) skew-normal (\citealt{azzalini_2013}; with skewness parameter $\alpha = 8$). In the case of the t-distribution, $\nu = 3$ is the smallest integer degrees of freedom resulting in $\sigma^2 < \infty$. Figure~\ref{fig:errors} in the Supplementary Material shows a comparison of the three different error distributions for the case where $\sigma^2 = 1$.

We consider ten different values for $n$: $20,40,\dots,200$. For the design points $\bx_1,\dots,\bx_n$, a requirement from establishing $\btheta_{L^2}$ as approximate the target parameter values (in Sections~\ref{sec:back} and~\ref{sec:methods}) was that the design points should be (quasi)-uniformly generated and be able to produce an approximately unbiased predictor $\hat{\mu}(\cdot; \by)$. A sensible candidate is a space-filling Latin hypercube design. In particular, we use maximum projection designs \citep{joseph_etal_2015}, implemented using the \texttt{R} package \texttt{MaxPro} \citep{maxpro}. 

Each of the four configurations have been considered previously by \cite{plumlee_2017}, \cite{wong_etal_2017}, \cite{gu_wang_2018}, and \cite{xie_xu_2020}. The comparisons considered here significantly expands on their treatment by studying long-run performance and a greater range of values of $n$. 

\paragraph{Configuration 1} 

The mathematical model, with $k=1$ and $p=2$, is
$$\eta(x, \btheta) = 7 \left[ \sin \left(2\pi \theta_1 - \pi \right) \right]^2 + 2 \left[ \left(2\pi \theta_2 - \pi \right)^2 \sin \left( 2 \pi x - \pi\right)\right],$$
where $\btheta = \left(\theta_1,\theta_2\right)^T$. The physical process is $\mu(x) = \eta(x, \btheta_0)$ with $\btheta_0 = \left(0.2,0.3\right)^T$, i.e. the mathematical model is exact. It follows that $\btheta_{L^2} = \btheta_0$, and we consider error variance $\sigma_0^2 = 0.2^2$. The prior distribution for $\btheta_{L^2}$ is such that the elements are independent with $\theta_{L^2,1} \sim \mathrm{U}(0,0.25)$ and $\theta_{L^2,2} \sim \mathrm{U}(0,0.5)$.

\paragraph{Configuration 2}
 
The mathematical model, with $k=1$ and $p=1$, is
$$\eta(x,\theta) = \sin \left( 5 \theta x\right) + 5x$$
and the physical process is
$$\mu(x) = 5x \cos \left( 15x/2\right) + 5x.$$
The value of $\theta$ minimizing $L_{L^2}(\theta)$ can be found numerically as $\theta_{L^2} = 1.8772$. We consider error variance $\sigma_0^2 = 0.2^2$. The prior distribution for $\theta_{L^2}$ is $\mathrm{U}\left(0,3\right)$.

\paragraph{Configuration 3}

The mathematical model, with $k=1$ and $p=1$, is
$$\eta(x,\theta) = \theta x$$
and the physical process is
$$\mu(x) = 4x + x \sin (5x).$$
The value of $\theta$ minimizing $L_{L^2}(\theta)$ is $\theta_{L^2} = 3.5653$. We consider error variance $\sigma_0^2 = 0.02^2$. The prior distribution for $\theta_{L^2}$ is $\mathrm{U}\left(2,5\right)$.

\paragraph{Configuration 4}

The mathematical model, with $k=2$ and $p=3$, is
$$\eta(\bx, \btheta) = 7 \sin^2 \left(2\pi \theta_1 - \pi\right) + 2 \left(2\pi\theta_2 - \pi\right)^2 \sin \left( 2\pi x_1 - \pi \right) + 6 \theta_3 \left(x_2 - \frac{1}{2}\right),$$
and the physical process is
$$\mu(\bx) = \eta(\bx, \btheta_0) + \cos \left(2\pi x_1 - \pi \right) + 2 \left(x_2^2 - x_2 + \frac{1}{6}\right),$$
where $\btheta_0  = \left(0.2,0.3,0.8\right)^T$. The values of $\btheta$ that minimise $L_{L^2}(\btheta)$ are $\btheta_{L^2} = \btheta_0$. We consider an error variance of (following \citealt{wong_etal_2017})
$$\sigma_0^2 = \frac{1}{10}\int_{\mathcal{X}} \left[ \mu(\bx) - \int_{\mathcal{X}} \mu(\bx)^2 \mathrm{d}\bx \right]^2 \mathrm{d}\bx = 0.7430.$$
We assume the following independent prior distributions for the elements of $\btheta_{L^2}$:
$$\theta_{L^2,1} \sim \mathrm{U}[0,0.25]; \qquad \theta_{L^2,2} \sim \mathrm{U}[0,0.5]; \qquad \theta_{L^2,3} \sim \mathrm{U}[0,1];$$
following the parameter spaces considered by \cite{wong_etal_2017}.

\subsection{Results}

We apply each of the seven methods to each configuration, each value of $n$ and each error distribution, $20,000$ times. For each application, we generate an MCMC sample of size 50,000. For the automatic bootstrapping approach, we use a bootstrap sample of $B=1,000$. We assess performance by mean (over the 20,000 repetitions) posterior mean, mean posterior standard deviation, and mean coverage of 95\% probability intervals for each of the $p$ calibration parameters.

There was negligible difference, for all values of $n$, between the results from the asymptotic and bootstrap specification of the scaling parameter $\gamma$. To aid in exposition, we have omitted the results from the bootstrap specification. Similarly, there was negligible difference between the results from the three different error distributions. We present results from the normal distribution here in the main manuscript and those from the t and skew-normal distributions in the Supplementary Material.

\begin{figure}
\includegraphics{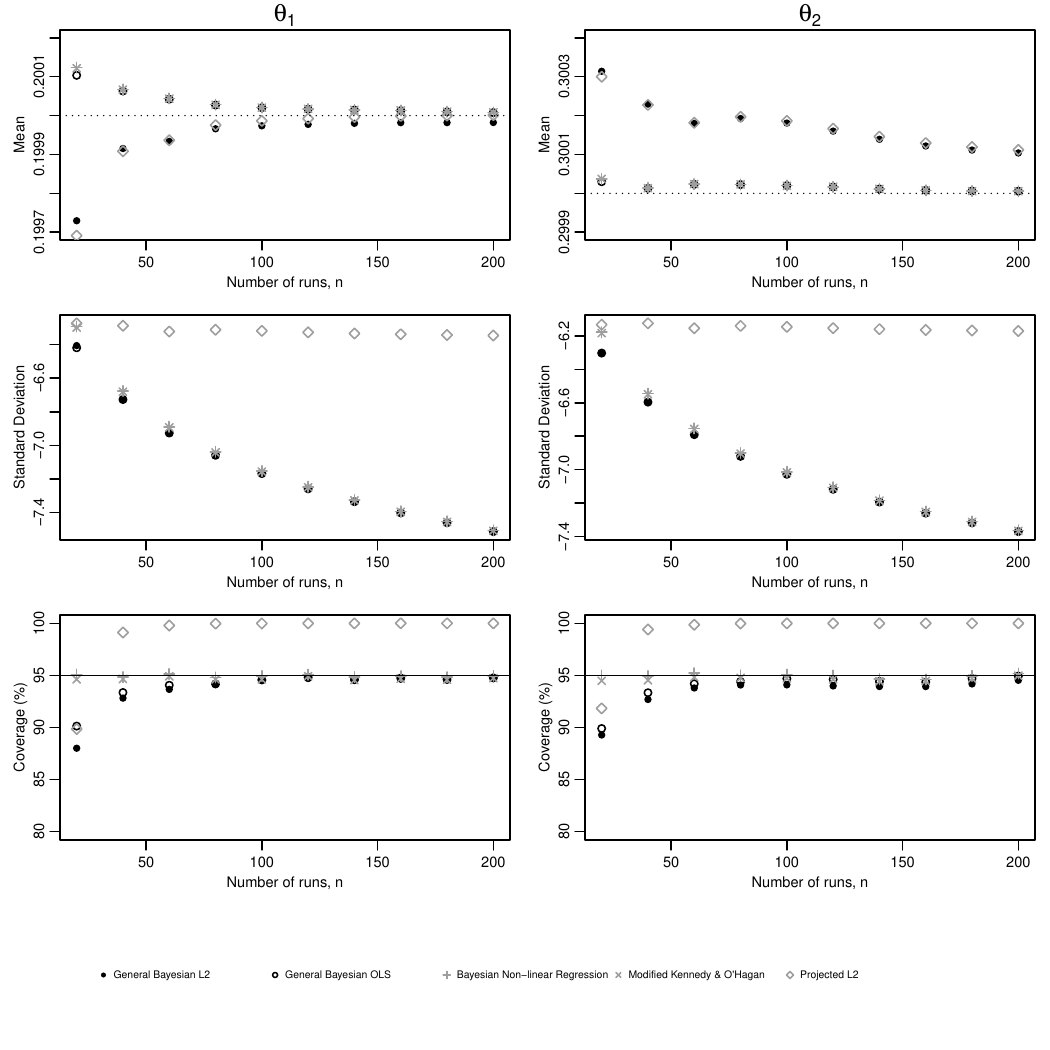}
\caption{For Configuration 1, plots showing the mean posterior mean,  log mean posterior standard deviation, and mean coverage of 95\% probability intervals for $\theta_1$ and $\theta_2$, for each of methods, for normally distributed errors. \label{fig:xu1}}
\end{figure}

\begin{figure}
\includegraphics{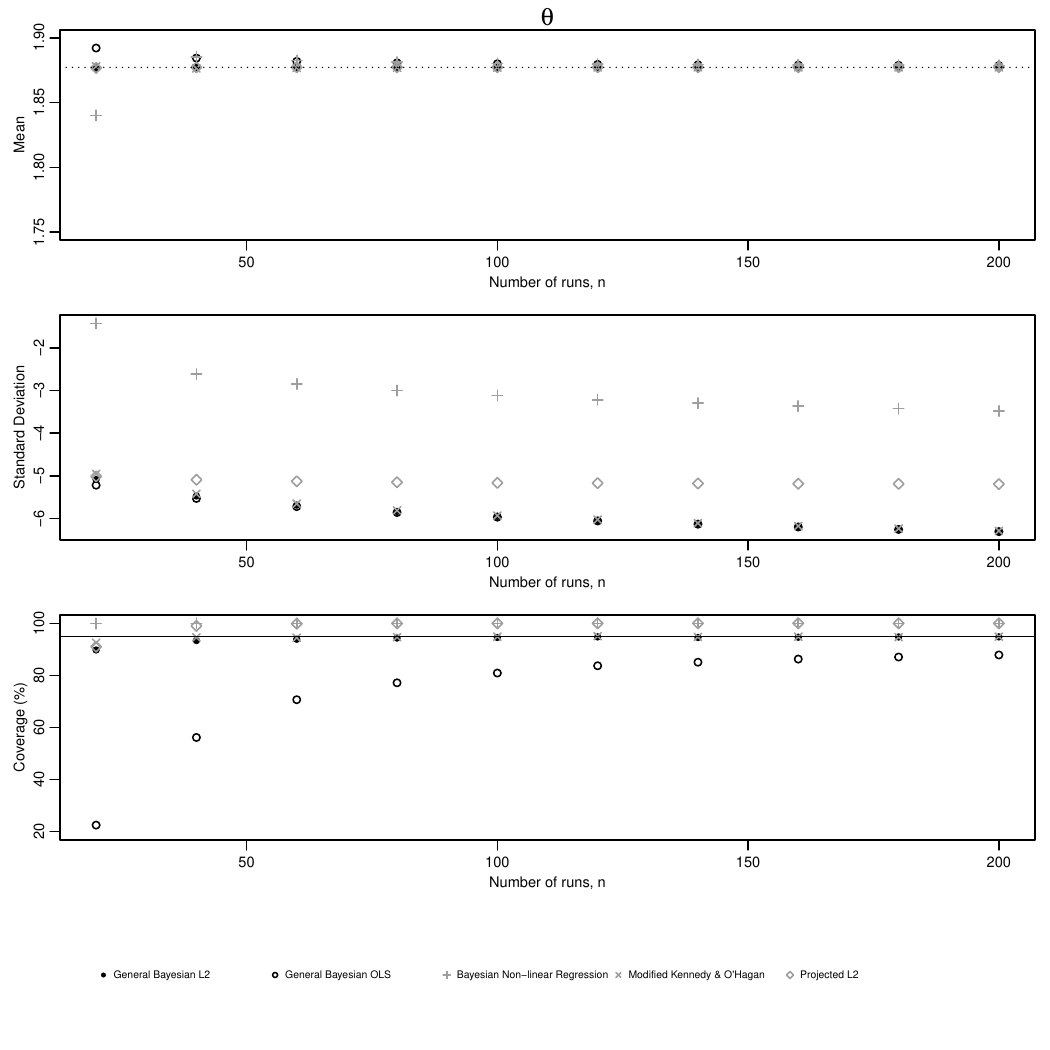}
\caption{For Configuration 2, plots showing the mean posterior mean, log mean posterior standard deviation, and mean coverage of 95\% probability intervals for $\theta$, for each of methods, for normally distributed errors. \label{fig:xu2}}
\end{figure}

\begin{figure}
\includegraphics{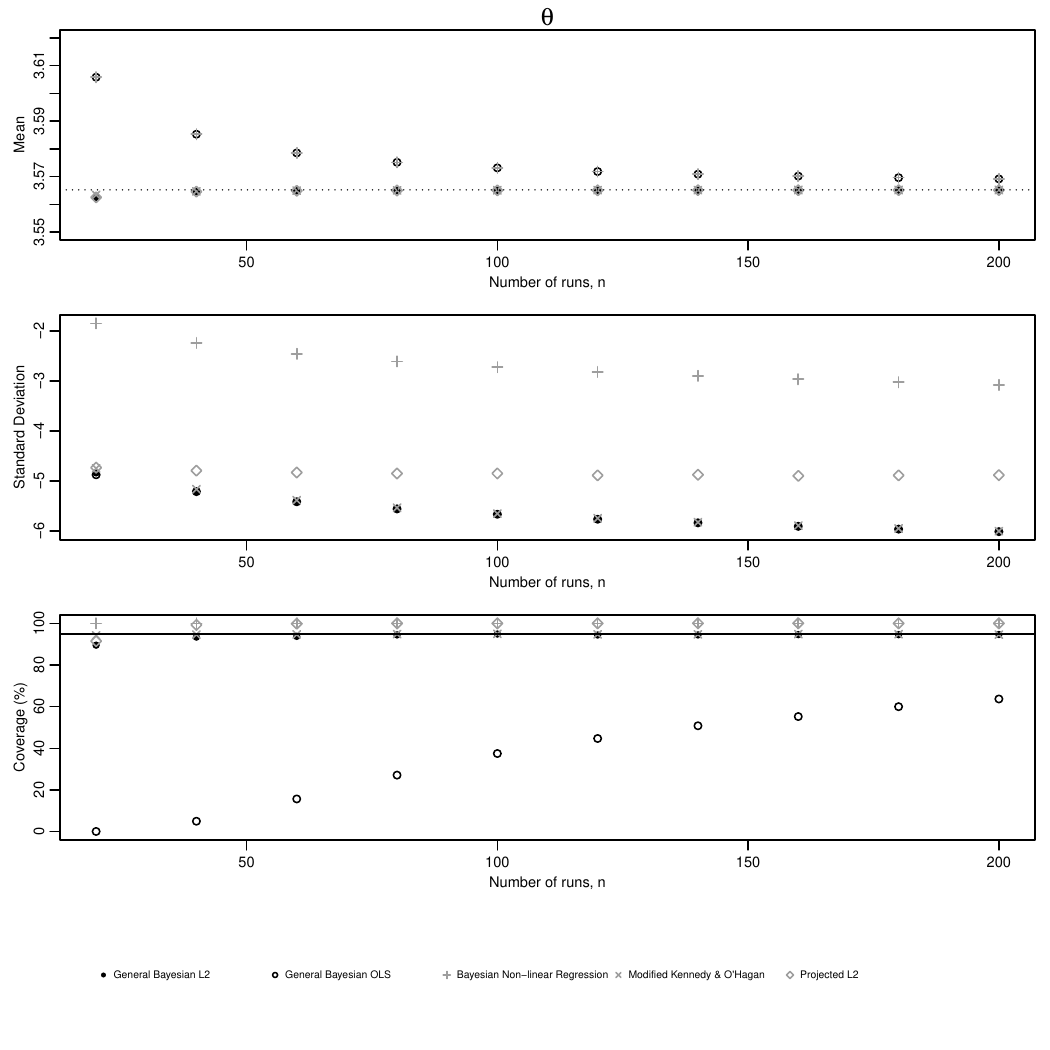}
\caption{For Configuration 3, plots showing the mean posterior mean,  log mean posterior standard deviation, and mean coverage of 95\% probability intervals for $\theta$, for each of methods, for normally distributed errors. \label{fig:xu3}}
\end{figure}

\begin{figure}
\includegraphics{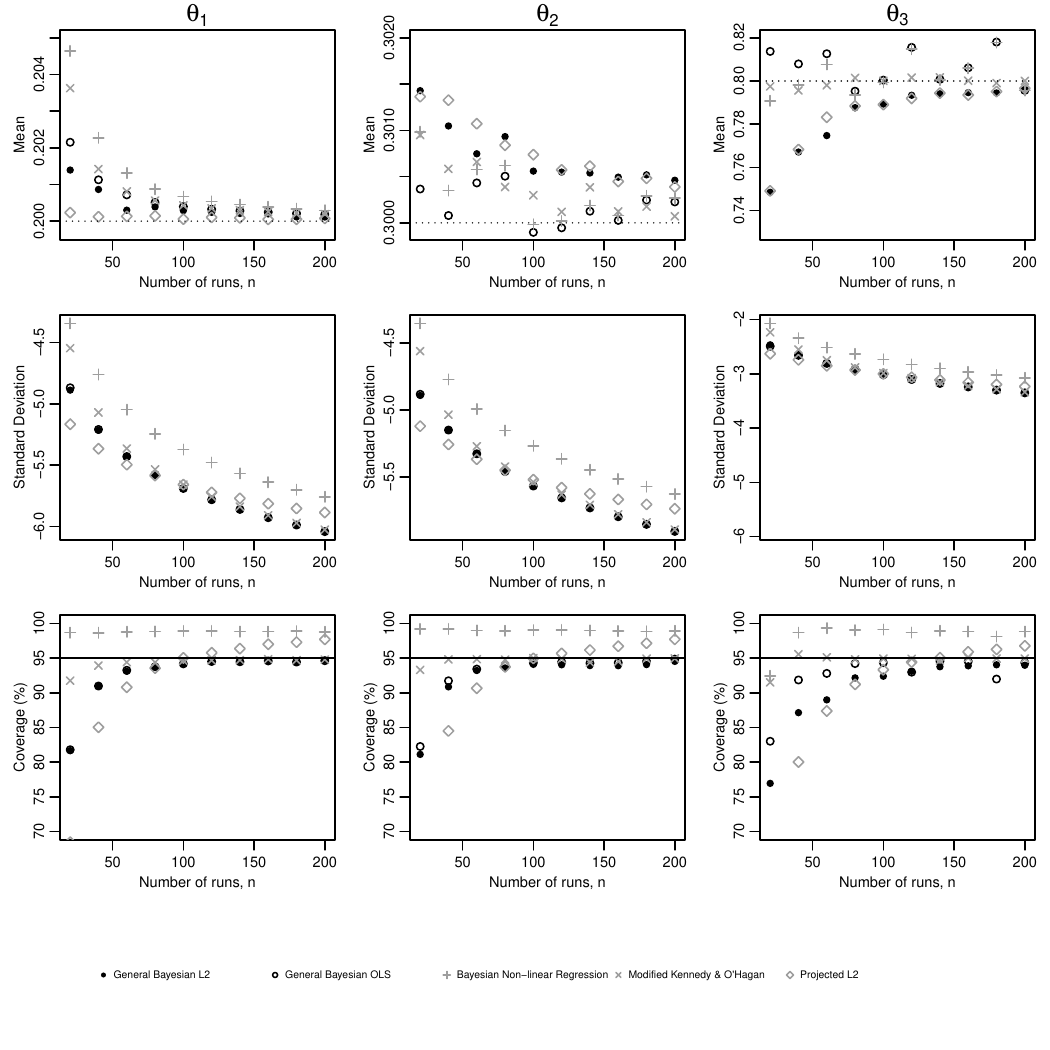}
\caption{For Configuration 4, plots showing the mean posterior mean,  log mean posterior standard deviation, and mean coverage of 95\% probability intervals for $\theta_1$, $\theta_2$, and $\theta_3$, for each of methods, for normally distributed errors. \label{fig:wong}}
\end{figure}

\begin{figure}
\includegraphics{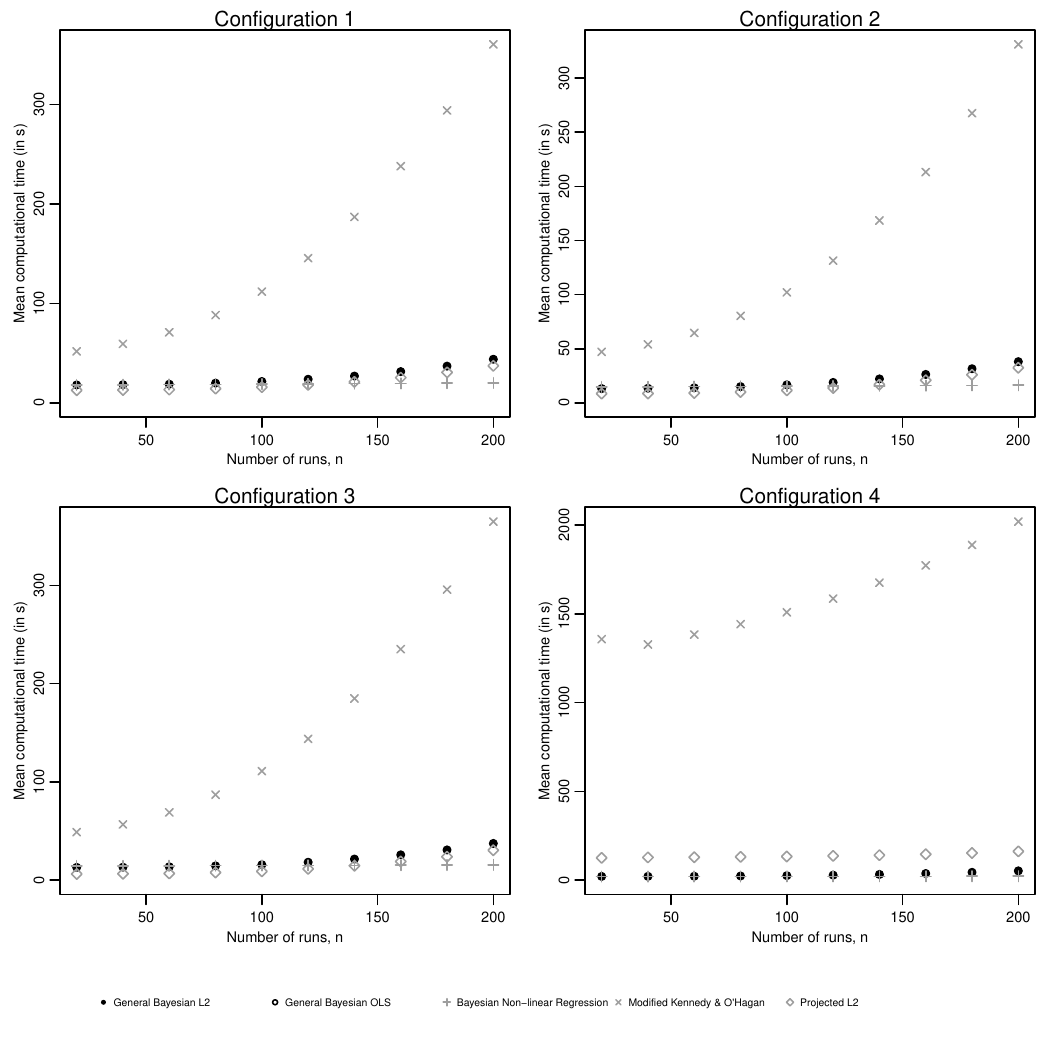}
\caption{Mean computational time (in s) against number of runs, $n$, for each of the five methods for normally distributed errors. Note that the timings for the general Bayesian $L^2$ and OLS approaches are indistinguishable. \label{fig:timings}}
\end{figure}

Figures~\ref{fig:xu1} to~\ref{fig:wong} show the results for Configurations 1 to 4, respectively, for normally distributed errors. The equivalent figures in the Supplementary Material for t distributed errors are Figures~\ref{fig:xu1_t} to~\ref{fig:wong_t}, and for skew-normally distributed errors are Figures~\ref{fig:xu1_sn} to~\ref{fig:wong_sn}. In each case, shown are the mean posterior mean (first row), log mean posterior standard deviation and mean coverage of 95\% probability intervals, against $n$, for each of the five methods. Each column of plots (where $p>1$) is for a different element of $\btheta$. Different plotting characters identify the different methods with black identifying the general Bayesian approaches ($L^2$ and OLS) and grey identifying standard Bayesian. 

The first rows of Figures~\ref{fig:xu1} to~\ref{fig:wong} show that the mean posterior mean converges towards $\btheta_{L^2}$ as $n$ increases. This confirms that, for these examples, the target parameters are approximately $\btheta_{L^2}$ for all five methods. However, there is significant noise for $\btheta_2$ in Configuration 4 (see Figure~\ref{fig:wong}). In terms of converging to the $\btheta_{L^2}$, no one method is uniformly superior. 

For Bayesian non-linear regression, the posterior standard deviation is largest and this inflation leads to over-coverage of the 95\% probability intervals. The exception to this is for Configuration 1 (where the mathematical model is exact) in Figure~\ref{fig:xu1}. This behaviour was predicted in Section~\ref{sec:bnlm}, where we found that the target parameter value for the response variance was greater than $\sigma_0^2$, with the difference being $\frac{1}{n}\sum_{i=1}^n \left[ \mu(\bx_i) - \eta(\bx_i;\btheta_{NLM})\right]^2$. This difference is a measure of the difference between the true physical system, $\mu(\cdot)$, and the mathematical model, $\eta(\cdot;\btheta_{NLM})$, at $\btheta_{NLM} \approx \btheta_{L^2}$. If the mathematical model is exact, then $\btheta_{L^2} = \btheta_0$, and the difference will be zero leading to no inflation of the posterior standard deviation. This is what is seen for Configuration 1 in Figure~\ref{fig:xu1}.

The projected $L^2$ calibration approach also appears to exhibit inflation of the posterior standard deviation leading to over-coverage of the probability intervals. However, unlike for Bayesian non-linear regression, it is not clear why this occurs.

For Configurations 2 and 3 (Figures~\ref{fig:xu2} and~\ref{fig:xu3}), general Bayesian OLS exhibits significant under-coverage of the probability intervals which does seem to correct as $n$ increases. This appears to be caused by the bias in estimating $\theta_{L^2}$ rather than a deflated posterior standard deviation since the posterior standard deviation under the general Bayesian OLS approach is very similar to as under the general Bayesian $L^2$ approach.

Discounting Configuration 1, when the mathematical model is exact, of the remaining two methods: general Bayesian $L^2$ and modified \citeauthor{kennedyohagan2001}, neither is uniformly superior. As an example, consider Configuration 4, with results shown in Figure~\ref{fig:wong}. The mean posterior mean for $\theta_1$ under general Bayesian $L^2$ is closer to $\theta_{L^2,1}$ than for the modified \citeauthor{kennedyohagan2001}, however, the reverse is true for $\theta_2$ and $\theta_3$. However, the coverage of the probability intervals under the modified \citeauthor{kennedyohagan2001} approach are closer to the nominal 95\%. These interpretations are typical for Configurations 2 and 3.

Figure~\ref{fig:timings} show the mean computational time (in s) against number of runs, $n$, for each of the five methods for normally distributed errors. Note that the timings for the general Bayesian $L^2$ and OLS approaches are indistinguishable. Clearly, the modified \citeauthor{kennedyohagan2001} approach is significantly more computationally expensive than the other approaches. This is due to the construction and inversion of the $n \times n$ covariance matrix that is required at each iteration of generating an MCMC sample.

\section{Real application: Wiffle balls} \label{sec:balls}

\begin{figure}
    \centering
    \includegraphics{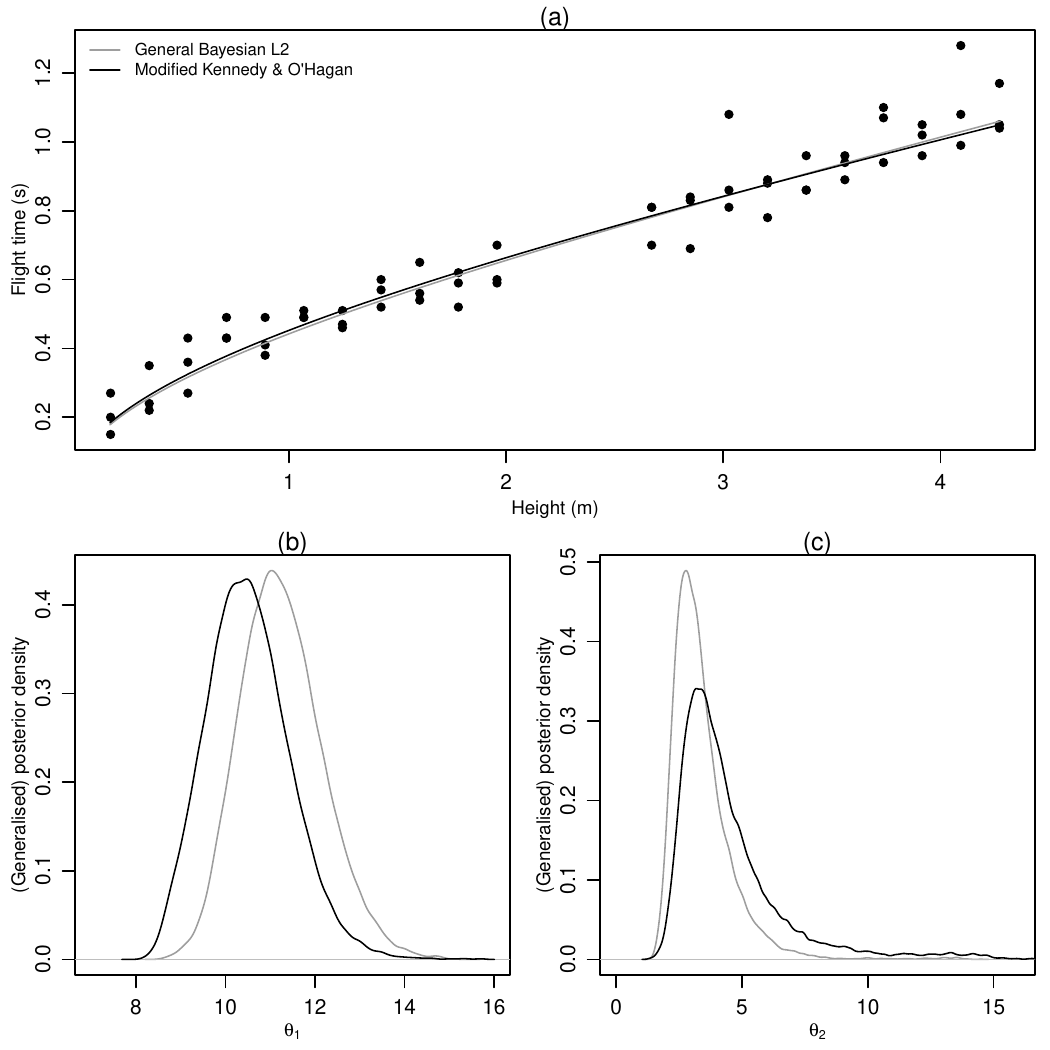}
    \caption{\emph{Results for the wiffle ball example. Panel (a) shows a plot of time against height. The lower panels show the generalized posterior densities of $\theta_1$ (b) and $\theta_2$ (c) for general Bayesian calibration under the $L^2$ loss and the modified \citeauthor{kennedyohagan2001} approaches. Panel (a) also shows $\eta(x, \tilde{\btheta})$ plotted against $x$ where $\tilde{\btheta}$ is the posterior mean from these two calibration methods.}}
    \label{fig:ball}
\end{figure}

The following example is described in \citet[][Section 8.1.2]{gramacy2020surrogates}. The experiment measured the time ($y$ in seconds) for a wiffle ball to freefall a height (in the range $[0.175,4.275]$ in metres, on the original scale). There are 3 replications at each of 21 different heights giving $n=63$. Figure~\ref{fig:ball}(a) shows a plot of the observed response, $y$, against height.

The mathematical model is derived incorporating non-linear air resistance (where the force due to air resistance is proportional to the square of the wiffle ball velocity) and is given by
$$\eta(x, \btheta) = \sqrt{\frac{\theta_2}{\theta_1}} \mathrm{arccosh}\left[ \exp \left( \frac{x}{\theta_2} \right)\right].$$
The parameter $\theta_1$ is nominally the acceleration due to gravity and $\theta_2$ is related to the air resistance and mass of the wiffle balls. The prior distribution for $\btheta_{L^2}$ is such that the elements are independent with $\theta_{L^2,1} \sim \mathrm{U}[0,20]$ and $\theta_{L^2,2} \sim \mathrm{U}[0,20]$.

Figure~\ref{fig:ball}(a) shows a plot of the observed values of time ($y_1,\dots,y_n$) against height. We apply general Bayesian $L^2$ and modified \citeauthor{kennedyohagan2001} calibration approaches. In each case, we generate an MCMC sample of size 250,000 from the (generalised) posterior distribution. 

Figures~\ref{fig:ball}(b) and (c) show plots of the posterior densities of $\theta_1$ and $\theta_2$, respectively, for the two different calibration approaches. There is significant agreement between the two approaches, with the conclusion that $\theta_{L^2} \approx (11, 3.5)$, for this particular $\eta(x, \btheta)$. To demonstrate the agreement, Figure~\ref{fig:ball}(a) shows $\eta(x, \tilde{\btheta})$ plotted against $x$ where $\tilde{\btheta}$ is the posterior mean from these two calibration methods. The two lines are almost indistinguishable.

\section{Discussion}

In this paper, we have proposed automatic scalings for general Bayesian calibration of mathematical models using $L^2$ and OLS losses. We then empirically compared these approaches to approaches in the literature that all approximately target the parameter values, $\btheta_{L^2}$, that minimise the squared difference between the true physical system and the mathematical model in the $L^2$ space. 

We found that all methods successfully target $\btheta_{L^2}$: mimicking the notion of unbiasedness. The difference between the approaches lay in their ability for uncertainty quantification. Bayesian non-linear regression and projected $L^2$ calibration resulted in over-coverage of probability intervals due to inflated posterior variance. By contrast, the general Bayesian OLS approach exhibited the opposite behaviour. 

We found that both the general Bayesian $L^2$ and the modified \citeauthor{kennedyohagan2001} \citep{plumlee_2017} approaches were both competitive. The general Bayesian $L^2$ approach assumes less about the data-generating process (e.g. errors are not assumed to be normally distributed). However, when we compared the methods using different heavy-tailed or skewed error distributions, this did not affect the results significantly. The modified \citeauthor{kennedyohagan2001} approach does carry significantly more computational expense.  

In the simulation studies and real application, the mathematical model, $\eta(\cdot, \btheta)$, is computationally inexpensive. If $\eta(\cdot, \btheta)$ is a computer model, then its evaluation can be replaced by the evaluation of the predictive mean of a surrogate model (e.g. a Gaussian process model). 

An issue for future research is the design of the experiment. In our empirical comparison, we used a space-filling design. This design was chosen to (approximately) ensure that the target parameters were $\btheta_{L^2}$. An alternative approach would be to use a decision-theoretic approach \cite{overstall_etal_2025} to sacrifice an increase in bias but gaining a reduction in posterior variance.

\setcounter{section}{0}
\setcounter{equation}{0}
\def\theequation{SM\arabic{section}.\arabic{equation}}
\def\thesection{SM\arabic{section}}
\def\thefigure{SM\arabic{figure}}
\def\thetable{SM\arabic{table}}

\section{Approximate target parameter values} \label{sec:targets}

\subsection{Bayesian non-linear regression} \label{sec:bnlm2}

The Kullback-Liebler divergence between the true probability distribution for the responses, and that imposed by the assumed model, is minimised by equivalently minimising
$$L_{NLM}(\btheta,\sigma^2) = \frac{n}{2} \log \sigma^2 + \frac{1}{2\sigma^2} \sum_{i=1}^n \left[ \mu(\bx_i) - \eta(\bx_i, \btheta\right]^2  + \frac{n \sigma_0^2}{\sigma^2},$$
with respect to $\btheta$ and $\sigma^2$. This is achieved when 
\begin{eqnarray}
\bzero_p & = & - \sum_{i=1}^n \frac{\partial \eta(\bx_i, \btheta_{NLM})}{\partial \btheta}  \left[ \mu(\bx_i) - \eta(\bx_i, \btheta_{NLM}\right]^2 \label{eqn:nlmtheta} \\
\lambda_{NLM} & = & \sigma_0^2 + \frac{1}{n} \sum_{i=1}^n \left[ \mu(\bx_i) - \eta(\bx_i, \btheta_{NLM}\right]^2. \label{eqn:nlmlambda2}
\end{eqnarray}
If the elements of the design $X = \left\{\bx_1,\dots,\bx_n\right\}$ is a uniform, or quasi-uniform, sample over $\mathcal{X}$, then $\btheta_{NLM}$ will approximate $\btheta_{L^2}$. To see this, the expectation of the right hand side of (\ref{eqn:nlmtheta}) under the distribution assumption for $X$ is
\begin{eqnarray*}
& & \mathrm{E}_X \left\{ - \sum_{i=1}^n \frac{\partial \eta(\bx_i, \btheta_{NLM})}{\partial \btheta}  \left[ \mu(\bx_i) - \eta(\bx_i, \btheta_{NLM}\right]^2 \right\} = \\
& & \qquad \qquad - \int_{\mathcal{X}} \frac{\partial \eta(\bx, \btheta_{NLM})}{\partial \btheta}  \left[ \mu(\bx) - \eta(\bx, \btheta_{NLM}\right]^2 \mathrm{d} \bx,
\end{eqnarray*}
with the right-hand side equal to $\bzero_p$ at $\btheta_{NLM} = \btheta_{L^2}$.

\subsection{Modified \citeauthor{kennedyohagan2001} calibration} \label{sec:mkoh2}

The target parameter values are given by minimising
\begin{eqnarray*}
L_{PKOH}(\btheta,\sigma^2, \kappa, \bpsi) & = & \frac{1}{2} \log \vert \Sigma_P (\btheta_{L^2}, \sigma^2, \kappa, \bpsi) \vert \\
& & \qquad + \frac{1}{2} \left[ \bmu_X - \bEta_X(\btheta)\right]^T \Sigma_P(\btheta_{L^2},\sigma^2, \kappa, \bpsi)^{-1} \left[ \bmu_X - \bEta_X(\btheta)\right]^T\\
& & \qquad \qquad + \frac{\sigma_0^2}{2}\mathrm{tr} \left[ \Sigma_P(\btheta_{L^2},\sigma^2, \kappa, \bpsi)^{-1} \right],
\end{eqnarray*}
with respect to $\btheta$, $\sigma^2$, $\kappa$ and $\bpsi$, where $\Sigma_P(\btheta,\sigma^2, \kappa, \bpsi) = \sigma^2 I_n + \sigma^2/\kappa C_{P,XX}(\btheta, \bpsi)$ with $C_{P,XX}(\btheta, \bpsi)$ being an $n \times n$ matrix with $ij$th element $c_P(\bx_i,\bx_j;\btheta,\bpsi)$. 

It can be shown that 
$$C_{P,XX}(\btheta, \bpsi) = C_{XX}(\bpsi) - C_{XD}(\bpsi)F_D(\btheta) \left[F_D(\btheta)^T C_{DD}(\bpsi) F_D(\btheta)\right]^{-1} F_D(\btheta)^TC_{XD}(\bpsi)^T,$$
where $C_{XD}(\bpsi)$ is an $n \times Q$ matrix with $iq$th element $c(\bx_i, \bchi_q;\bpsi)$, $C_{DD}(\bpsi)$ is an $Q \times Q$ matrix with $iq$th element $c(\bchi_i, \bchi_q;\bpsi)$, and $F_D(\btheta)$ is a $Q \times p$ matrix with $q$th row $\omega_q \partial \eta(\bchi_q,\btheta)/\partial \btheta^T$. 

The gradient of $L_{PKOH}(\btheta,\sigma^2, \kappa, \bpsi)$ with respect to $\btheta$ is 
$$\frac{ \partial L_{PKOH}(\btheta,\sigma^2, \kappa, \bpsi)}{\partial \btheta} = - \frac{\partial \bEta_X(\btheta)}{\partial \btheta} \Sigma_P(\btheta_{L^2},\sigma^2, \kappa, \bpsi)^{-1} \left[ \bmu_X - \bEta_X(\btheta) \right],$$ 
where $\partial \bEta_X(\btheta)/\partial \btheta$ is the $p \times n$ Jacobian of $\bEta_X(\btheta)$, with $ji$th element $\partial \eta(\bx_i,\btheta)/\partial \theta_j$, and 
$$\Sigma_P(\btheta,\sigma^2, \kappa, \bpsi)^{-1} = \frac{1}{\sigma^2}I_n - \frac{1}{\sigma^2} \left[\kappa C_{P,XX}(\btheta, \bpsi)^{-1} + I_n \right]^{-1},$$ 
which follows from the Woodbury matrix identity \citep[e.g.][]{hendersonsearle_1981}. Further application of the Woodbury matrix identity gives
\begin{eqnarray*}
C_{P,XX}(\btheta, \bpsi)^{-1}  &=& \\
C_{XX}(\bpsi)^{-1}&+& C_{XX}(\bpsi)^{-1}C_{XD}(\bpsi)F_D(\btheta) \left\{ F_D(\btheta)^T U F_D(\btheta) \right\}^{-1} F_D(\btheta)^TC_{XD}(\bpsi)^TC_{XX}(\bpsi)^{-1},
\end{eqnarray*}
where $U = \left[ C_{DD}(\bpsi) - C_{XD}(\bpsi)^T C_{XX}(\bpsi)^{-1}C_{XD}(\bpsi)\right]$.

The finite version of $\bzero_p = -2 \int_{\mathcal{X}} \frac{\partial \eta(\bx; \btheta_{L^2})}{\partial \btheta} \delta_{L^2}(\bx) \mathrm{d}\bx$ is $\bzero_p = -2 F_D(\btheta_{L^2})^T \bdelta_{D:L^2}$, where $\bdelta_{D:L^2} = \left[ \delta_{L^2}(\bchi_1), \dots, \delta_{L^2}(\bchi_Q)\right]^T$. Let $\bdelta_{X:L^2} = \left[ \delta_{L^2}(\bx_1), \dots, \delta_{L^2}(\bx_n)\right]^T$ and assume $\delta_{L^2}(\bx)$ is a zero-mean Gaussian process with covariance function $\gamma c(\cdot,\cdot; \bpsi)$. Then
$$\bdelta_{D:L^2} \vert \bdelta_{X:L^2} \sim \mathrm{N} \left\{ C_{XD}(\bpsi)^T C_{XX}(\bpsi)^{-1} \bdelta_{X:L^2}, \gamma \left[ C_{DD}(\bpsi) - C_{XD}(\bpsi)^T C_{XX}(\bpsi)^{-1}C_{XD}(\bpsi) \right] \right\}.$$
If $\bzero_p = -2 F_D(\btheta_{L^2})^T \bdelta_{D:L^2}$, then
\begin{equation}
\begin{array}{lclcl}
\bzero_p & = & \mathrm{E}_{\bdelta_{D:L^2} \vert \bdelta_{X:L^2}} \left[ -2 F_D(\btheta_{L^2})^T \bdelta_{D:L^2} \right] & = & -2 F_D(\btheta_{L^2})^T C_{XD}(\bpsi)^T C_{XX}(\bpsi)^{-1} \bdelta_{X:L^2},\\
\bzero_{p \times p} & = & \mathrm{var}_{\bdelta_{D:L^2} \vert \bdelta_{X:L^2}} \left[ -2 F_D(\btheta_{L^2})^T \bdelta_{D:L^2} \right] & = & 4 \gamma F_D(\btheta_{L^2})^T U F_D(\btheta_{L^2}).
\end{array}
\label{eqn:constraints}
\end{equation}
Then as $\btheta \to \btheta_{L^2}$, from (\ref{eqn:constraints}), $\Sigma_P(\btheta,\sigma^2, \kappa, \bpsi)^{-1} \left[ \bmu_X - \bEta_X(\btheta) \right] \to \frac{1}{\sigma^2} \left[ \bmu_X - \bEta_X(\btheta) \right]$, meaning that
$$\frac{ \partial L_{PKOH}(\btheta_{L^2},\sigma^2, \kappa, \bpsi)}{\partial \btheta} \to - \frac{1}{\sigma^2}\frac{\partial \bEta_X(\btheta_{L^2})}{\partial \btheta}\left[ \bmu_X - \bEta_X(\btheta_{L^2}) \right] \approx \bzero_p$$
for a uniform (or quasi-uniform) design $X$. Therefore, the target calibration parameters, $\btheta_{PKOH}$, are approximately $\btheta_{L^2}$.

\subsection{Projected $L^2$ calibration} \label{sec:xx2}

Consider the target parameter values, $\btheta_{PL}$. Given target values for $\bmu_D = \left[ \mu(\bchi_1),\dots,\mu(\bchi_Q) \right]^T$, the target calibration parameters are given by minimising (\ref{eqn:xiexi}). For determining the target values, $\bmu_{D,PL}$, for $\bmu_D$, the assumptions of the Gaussian process model imply that $\by \vert \bmu_D \sim \mathrm{N} \left[ C_{XD}(\bpsi)C_{DD}(\bpsi)^{-1} \bmu_D, \gamma \left\{ \Phi(\bpsi,\kappa) - C_{XD}(\bpsi)C_{DD}(\bpsi,\kappa)^{-1}C_{XD}(\bpsi)^T\right\} \right]$. The resulting log-likelihood tells us that the target values for $\bmu_D$ are given by minimising
\begin{eqnarray}
L_{PL}(\bmu_D) & = & \frac{1}{2\gamma} \left[ \bmu_X - C_{XD}(\bpsi)C_{DD}(\bpsi,\kappa)^{-1} \bmu_D\right]^T \nonumber \\
& & \qquad \times \left\{ \Phi(\bpsi,\kappa) - C_{X,D}(\bpsi)C_{DD}(\bpsi,\kappa)^{-1}C_{X,D}(\bpsi)^T\right\}^{-1}\nonumber \\
& & \qquad \qquad \times \left[ \bmu_X - C_{X,D}(\bpsi)C_{DD}(\bpsi,\kappa))^{-1} \bmu_D\right],
\label{eqn:lossXX}
\end{eqnarray}
with respect to $\bmu_D$. It can be shown that the target values, $\bmu_{D,PL}$ will solve the equation $\bmu_X = C_{XD}(\bpsi)C_{DD}(\bpsi,\kappa)^{-1} \bmu_{D,PL}$. Therefore,
$$\btheta_{PL} = \arg \min_{\btheta} \sum_{q=1}^Q \omega_q \left[\mu_{D,PL,q} - \eta(\btheta,\bchi_q) \right]^2,$$
where $\mu_{D,PL,q}$ is the $q$th element of $\bmu_{D,PL}$. Using properties of pseudo-inverses \citep[e.g.][Section 3.6.3]{gentle2007},
$$\bmu_{D,PL} = C_{DD}(\bpsi,\kappa) \left[ C_{X,D}(\bpsi)^T C_{XD}(\bpsi) \right]^{-1} C_{X,D}(\bpsi)^T \bmu_X.$$
If $\mu(\cdot)$ is assumed to have a Gaussian process with correlation function $\gamma c(\cdot,\cdot;\bpsi)$, then the expectation of $\bmu_{PL}$ (conditional on $\bmu_D$) is equal to $\bmu_D$. Therefore, $\bmu_{PL}$ is approximately $\bmu_D$, and (\ref{eqn:xiexi}) is minimised by $\btheta_{PL} = \btheta_{L^2}$.

\section{Automatic scaling using bootstrapping} \label{sec:automaticboot}

\subsection{General Bayesian $L^2$ calibration}

Define the $i$th residual as $\hat{\epsilon}_i = y_i - \hat{\mu}(\bx_i;\by)$, for $i=1,\dots,n$. Then the following steps give the bootstrap specification of $\gamma$.

\begin{enumerate}
\item
For $b=1,\dots,B$, complete the following steps.
\begin{enumerate}
\item
Sample, with replacement, $n$ values denoted $\hat{\epsilon}_1^{(b)}, \dots, \hat{\epsilon}_n^{(b)}$ from the residuals $\hat{\epsilon}_1,\dots,\hat{\epsilon}_n$.
\item
Set $y_i^{(b)} = \hat{\mu}(\bx_i;\by) + \hat{\epsilon}_i^{(b)}$, for $i=1,\dots,n$.
\item \label{step:2c}
Find $\hat{\btheta}_{L^2}^{(b)} = \arg \min_{\btheta} \ell_{L^2}(\btheta;\by^{(b)})$, where $\by^{(b)} = \left(y_1^{(b)},\dots,y_n^{(b)}\right)^T$, and set
$$\Lambda_0^{(b)} = 2 \left[ \ell_{L^2}(\hat{\btheta}_{L^2};\by^{(b)}) - \ell_{L^2}(\hat{\btheta}^{(b)}_{L^2};\by^{(b)})\right].$$
\end{enumerate}
\item
Set 
$$\gamma = \frac{p}{\frac{1}{B}\sum_{b=1}^B \Lambda_0^{(b)}}.$$
\end{enumerate}

\subsection{General Bayesian OLS calibration}

A bootstrap approach can be adopted to specify $\gamma$. This follows the algorithm in Section~\ref{sec:automaticboot}, but replacing Step~\ref{step:2c} by the following.

\begin{enumerate}
\item[(c)]
Find $\hat{\btheta}_{OLS}^{(b)} = \arg \min_{\btheta} \ell_{OLS}(\btheta;\by^{(b)})$, where $\by^{(b)} = \left(y_1^{(b)},\dots,y_n^{(b)}\right)^T$, and set
$$\Lambda_0^{(b)} = 2 \left[ \ell_{OLS}(\hat{\btheta}_{OLS};\by^{(b)}) - \ell_{OLS}(\hat{\btheta}^{(b)}_{OLS};\by^{(b)})\right].$$
\end{enumerate}

\section{Error distributions} \label{sec:errors}

\begin{figure}
\includegraphics{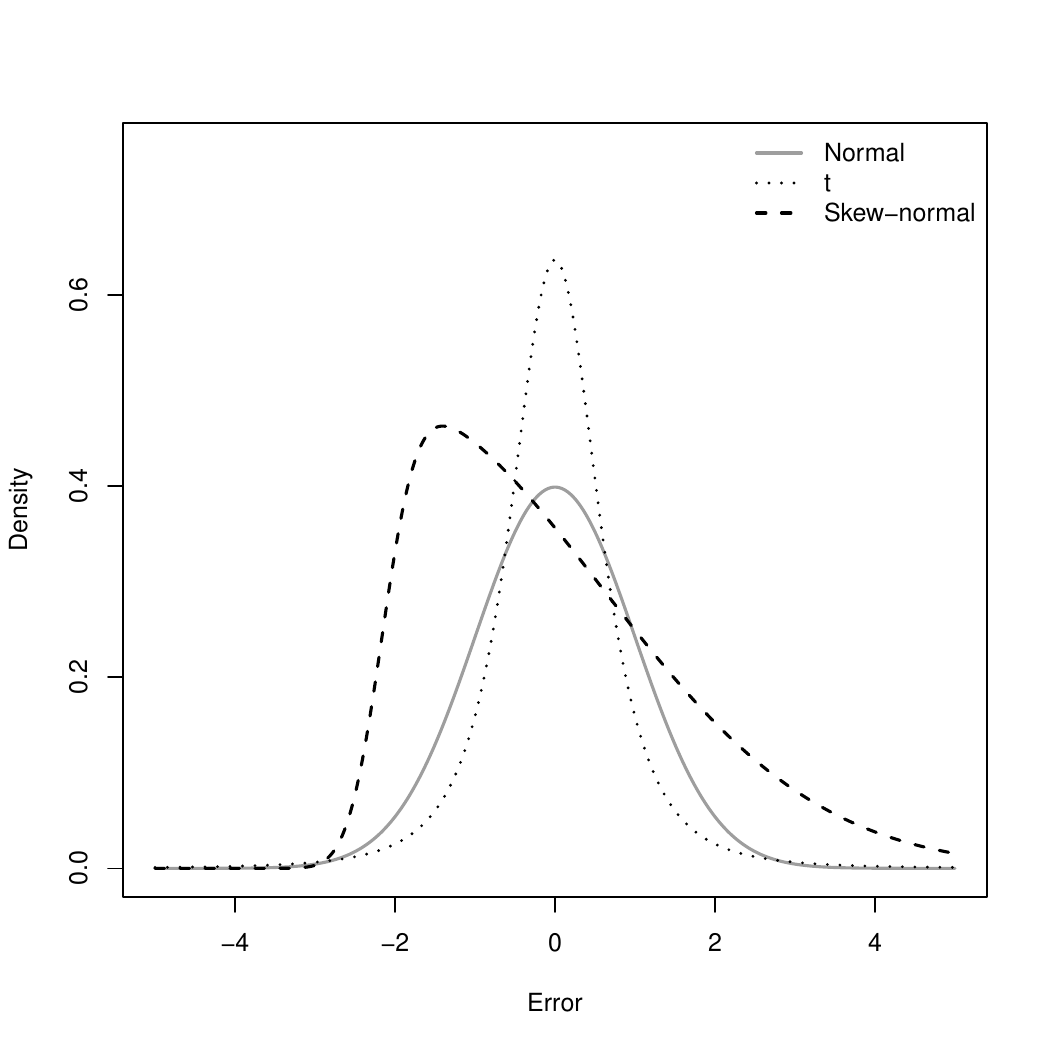}
\caption{A comparison of the three different error distributions. In each case, the mean and variance are 0 and 1 respectively. The t-distribution has $\nu = 3$ degrees of freedom and the skew-normal has a skewness parameter of $\alpha = 8$. \label{fig:errors}}
\end{figure}

\section{Additional results} \label{sec:addres}

\begin{figure}
\includegraphics{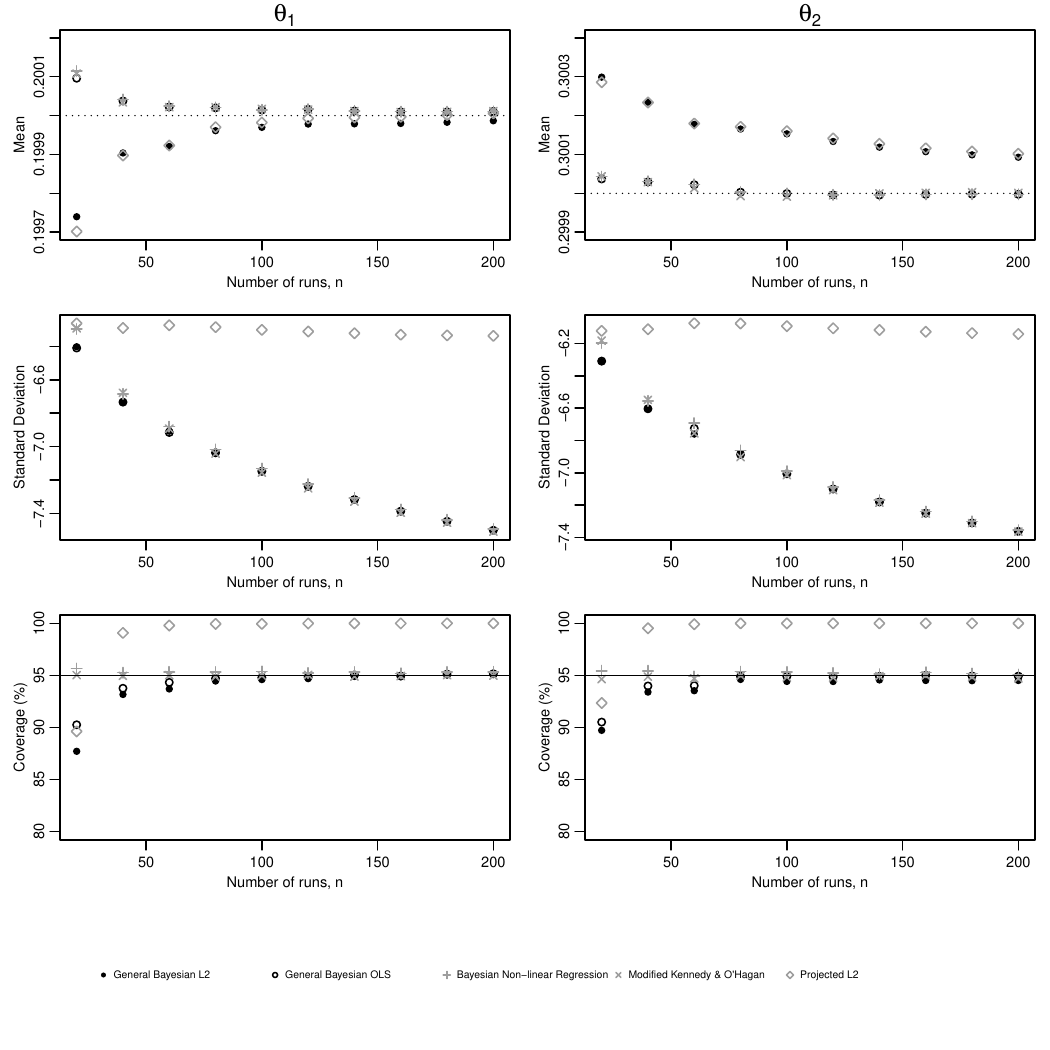}
\caption{For Configuration 1, plots showing the mean posterior mean,  log mean posterior standard deviation, and mean coverage of 95\% probability intervals for $\theta_1$ and $\theta_2$, for each of methods, for t-distributed errors. \label{fig:xu1_t}}
\end{figure}

\begin{figure}
\includegraphics{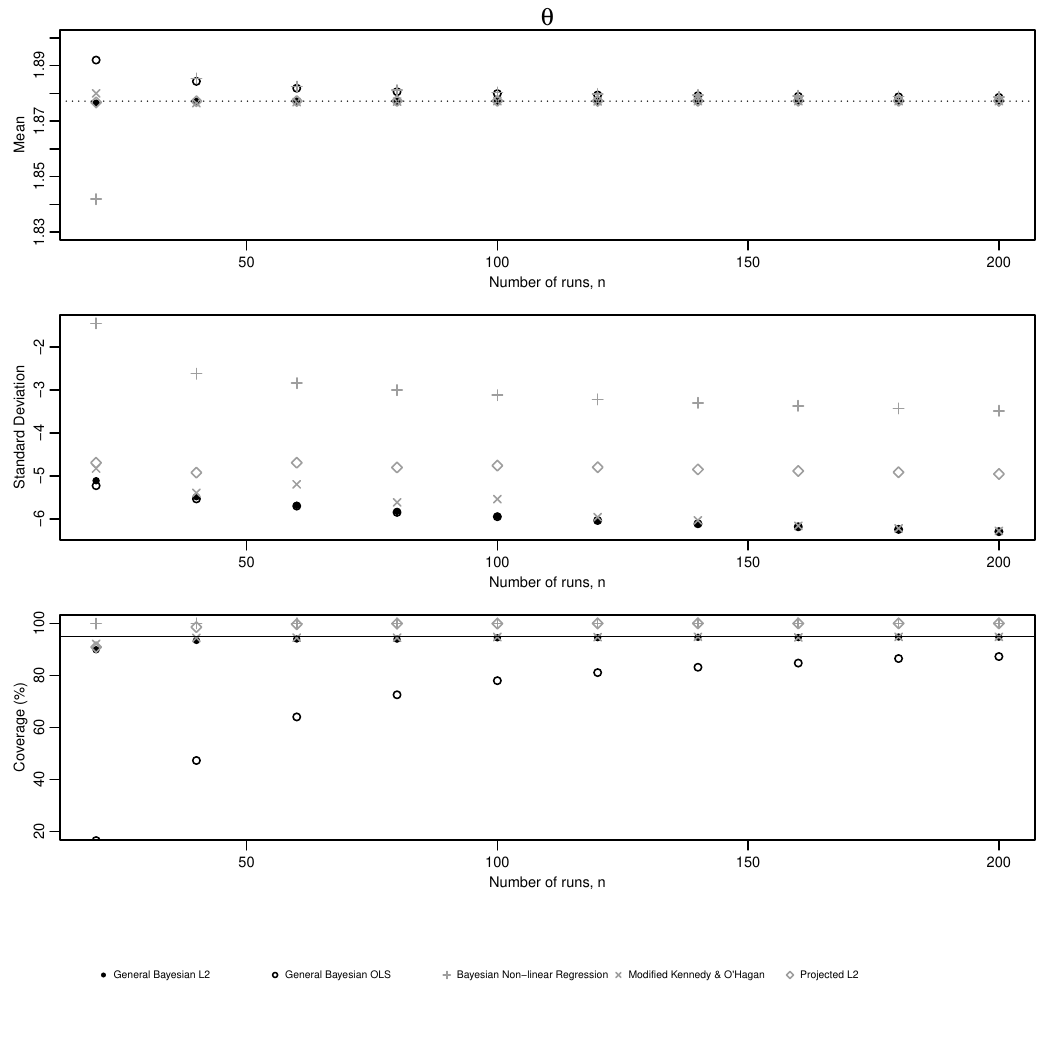}
\caption{For Configuration 2, plots showing the mean posterior mean, log mean posterior standard deviation, and mean coverage of 95\% probability intervals for $\theta$, for each of methods, for t-distributed errors. \label{fig:xu2_t}}
\end{figure}

\begin{figure}
\includegraphics{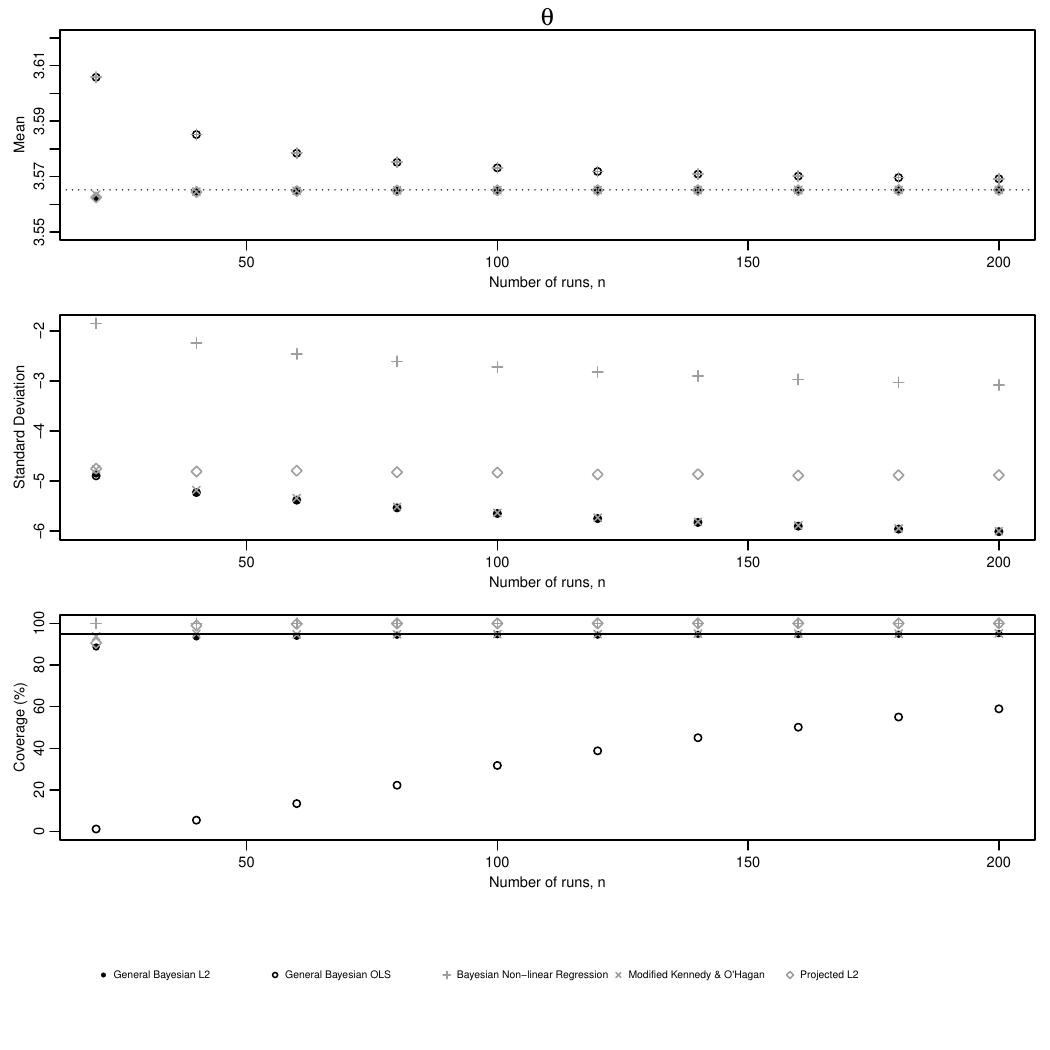}
\caption{For Configuration 3, plots showing the mean posterior mean,  log mean posterior standard deviation, and mean coverage of 95\% probability intervals for $\theta$, for each of methods, for t-distributed errors. \label{fig:xu3_t}}
\end{figure}

\begin{figure}
\includegraphics{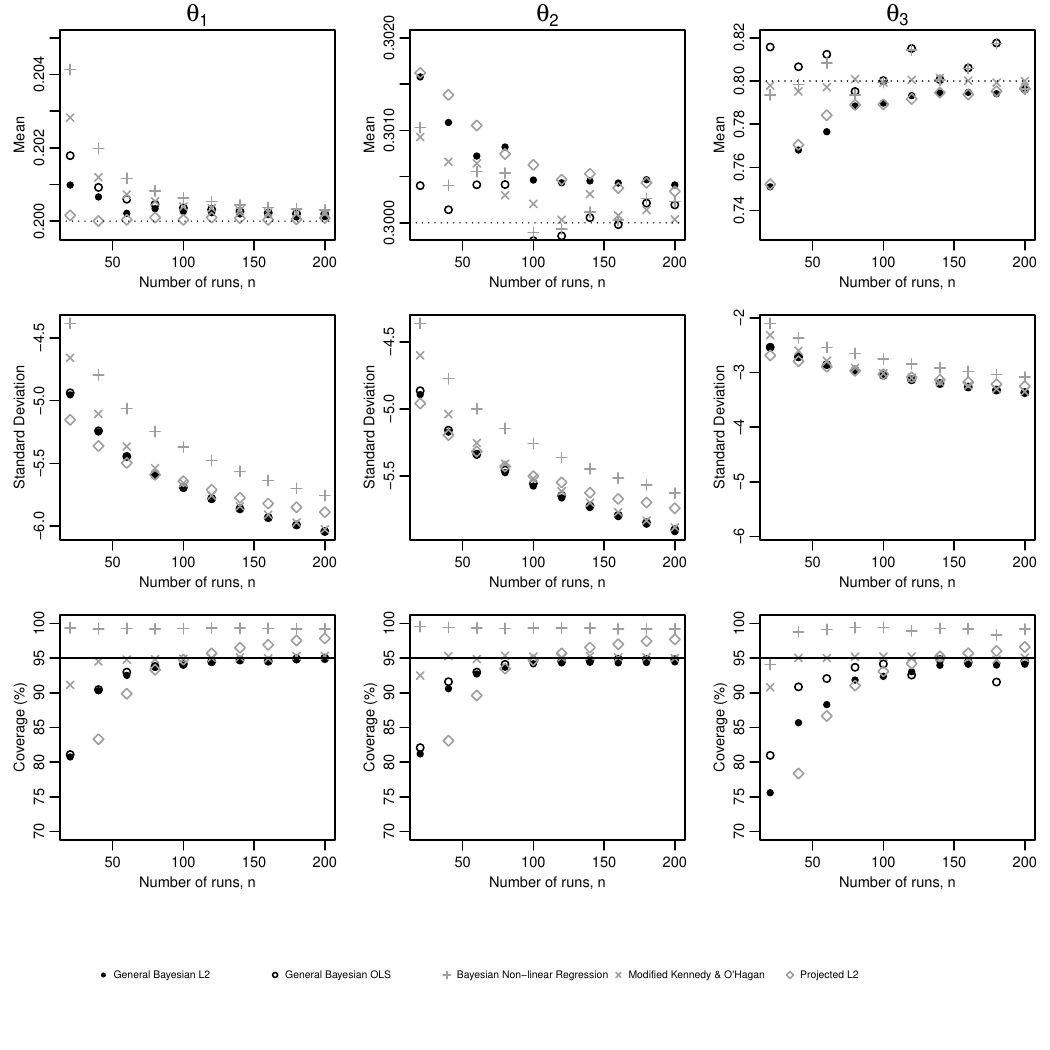}
\caption{For Configuration 4, plots showing the mean posterior mean,  log mean posterior standard deviation, and mean coverage of 95\% probability intervals for $\theta_1$, $\theta_2$, and $\theta_3$, for each of methods, for t-distributed errors. \label{fig:wong_t}}
\end{figure}

\begin{figure}
\includegraphics{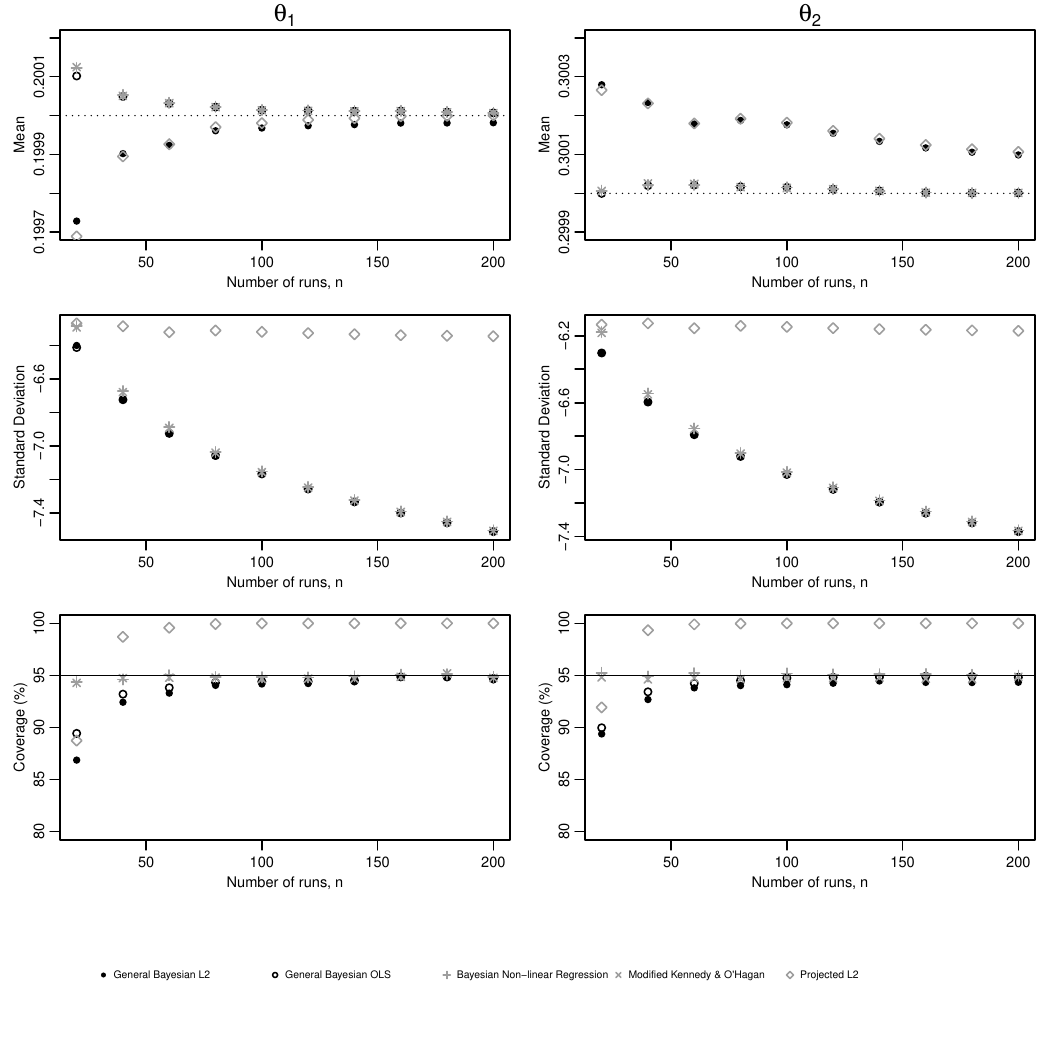}
\caption{For Configuration 1, plots showing the mean posterior mean,  log mean posterior standard deviation, and mean coverage of 95\% probability intervals for $\theta_1$ and $\theta_2$, for each of methods, for skew normally distributed errors. \label{fig:xu1_sn}}
\end{figure}

\begin{figure}
\includegraphics{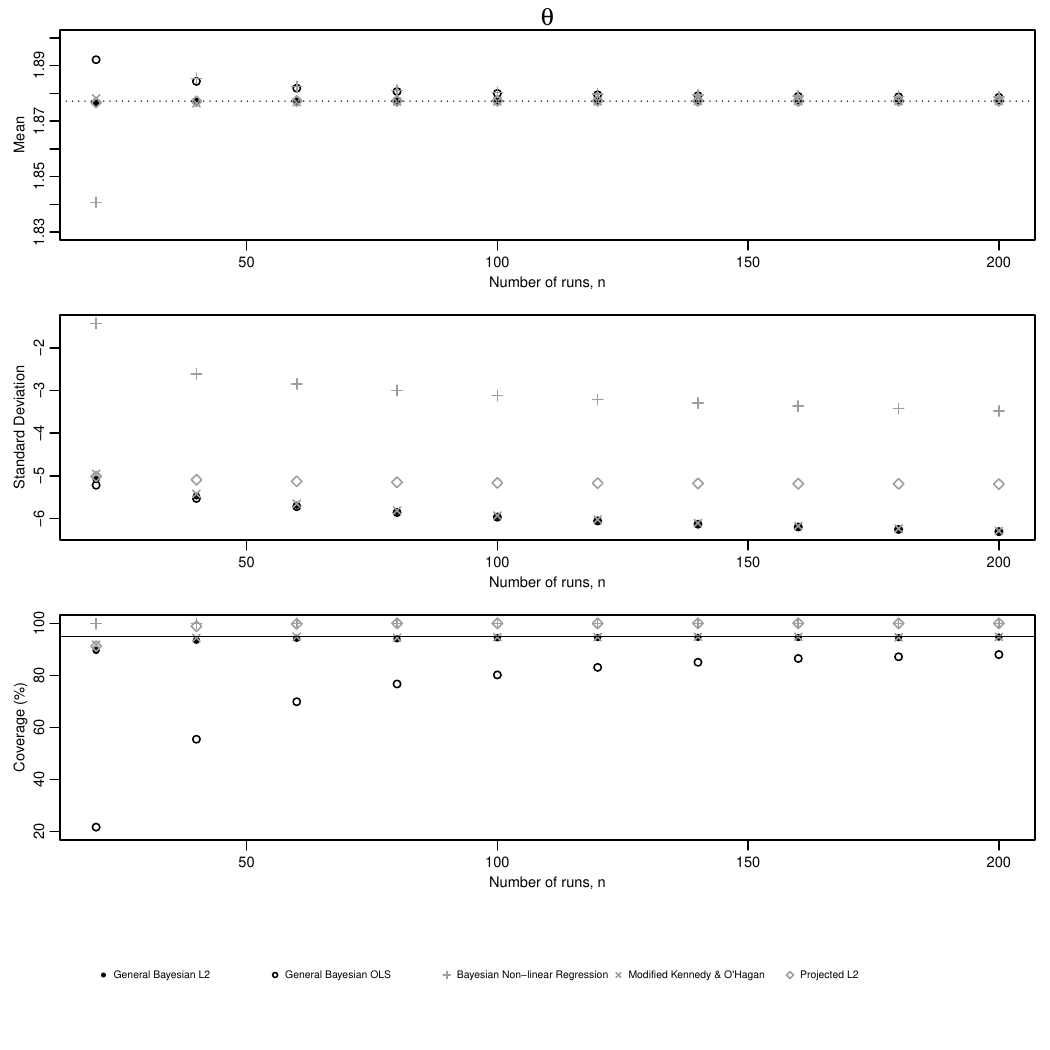}
\caption{For Configuration 2, plots showing the mean posterior mean, log mean posterior standard deviation, and mean coverage of 95\% probability intervals for $\theta$, for each of methods, for skew normally distributed errors. \label{fig:xu2_sn}}
\end{figure}

\begin{figure}
\includegraphics{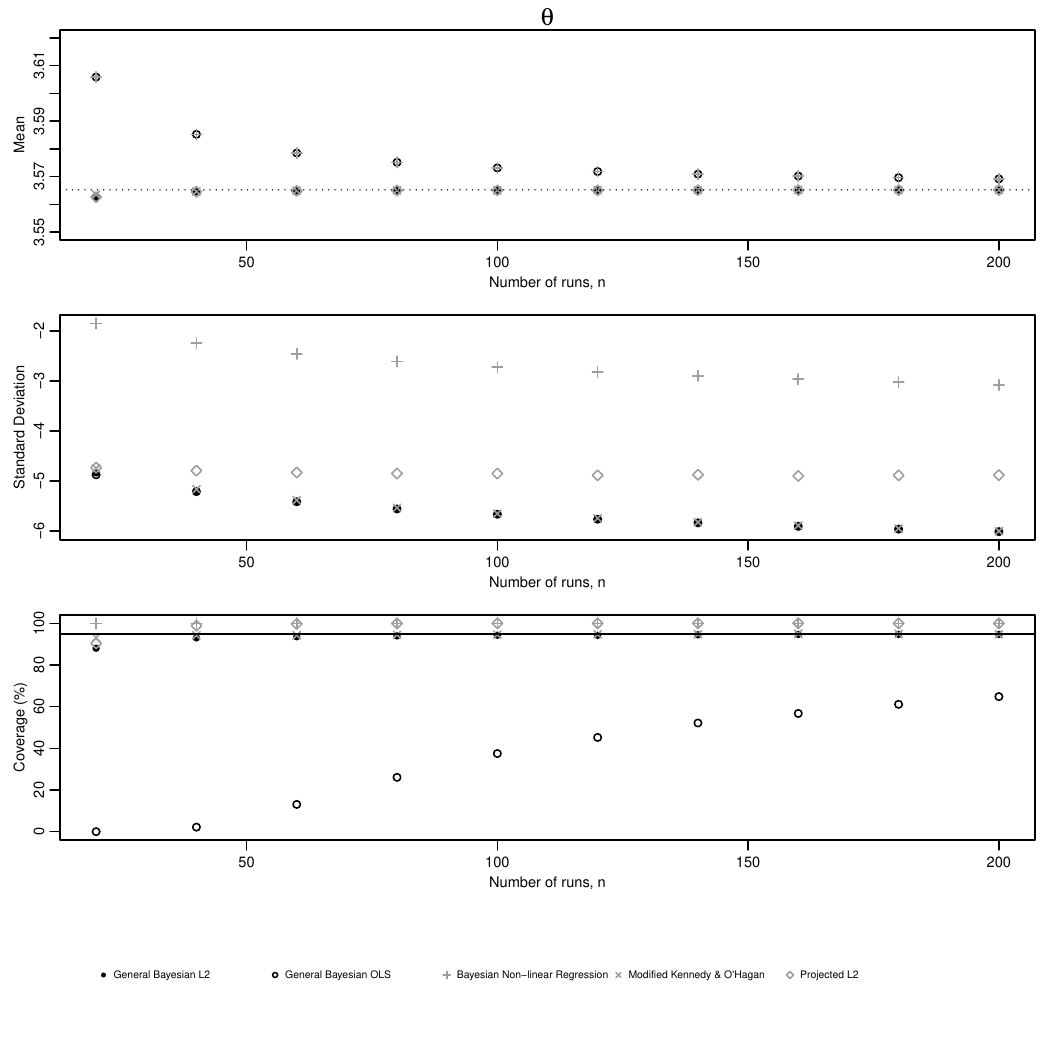}
\caption{For Configuration 3, plots showing the mean posterior mean,  log mean posterior standard deviation, and mean coverage of 95\% probability intervals for $\theta$, for each of methods, for skew normally distributed errors. \label{fig:xu3_sn}}
\end{figure}

\begin{figure}
\includegraphics{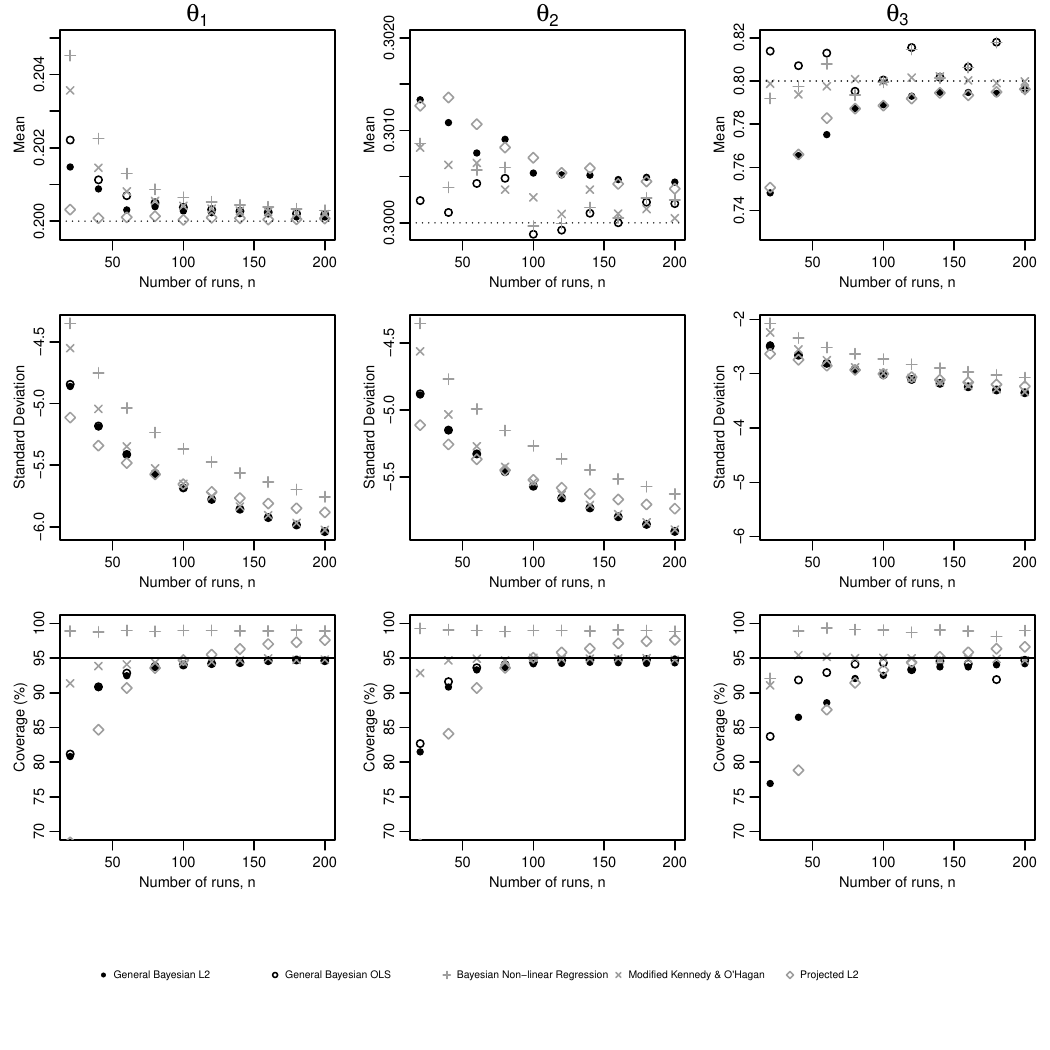}
\caption{For Configuration 4, plots showing the mean posterior mean,  log mean posterior standard deviation, and mean coverage of 95\% probability intervals for $\theta_1$, $\theta_2$, and $\theta_3$, for each of methods, for skew normally distributed errors. \label{fig:wong_sn}}
\end{figure}

\bibliographystyle{rss}

\bibliography{calib}
\end{document}